\documentclass[aps,superscriptaddress,amsmath,amssymb,floatfix,twocolumn,showpacs,amsfonts,longbibliography]{revtex4-2}

\usepackage{graphicx}
\usepackage{dcolumn}
\usepackage{bm}
\usepackage{physics}
\usepackage{times}
\usepackage[varg]{txfonts}
\usepackage{textcomp}
\usepackage{graphicx}
\usepackage{subfigure}
\usepackage{tabu}
\usepackage{color}
\usepackage[colorlinks=true,citecolor=blue,urlcolor=blue,linkcolor=blue,hyperindex]{hyperref}
\usepackage{braket}
\usepackage{overpic}
\usepackage{amssymb}
\usepackage{multirow}
\usepackage{verbatim}
\title{My Title Here}
\usepackage[]{placeins}
\usepackage{ulem}

\usepackage[usenames,dvipsnames]{xcolor}
\usepackage{tcolorbox}
\usepackage{tabularx}
\usepackage{array}
\usepackage{colortbl}
\tcbuselibrary{skins}
\usepackage{multirow}

\newcolumntype{Y}{>{\raggedleft\arraybackslash}X}

\tcbset{tab1/.style={fonttitle=\bfseries\large,fontupper=\normalsize\sffamily,
colback=yellow!10!white,colframe=red!75!black,colbacktitle=Salmon!40!white,
coltitle=black,center title,freelance,frame code={
\foreach \n in {north east,north west,south east,south west}
{\path [fill=red!75!black] (interior.\n) circle (3mm); };},}}

\tcbset{tab2/.style={enhanced,fonttitle=\bfseries,fontupper=\normalsize\sffamily,
colback=white,colframe=blue!0!black,colbacktitle=Salmon!40!white,
coltitle=black,center title}}

\usepackage{mathtools}
\usepackage{siunitx}
\DeclarePairedDelimiterXPP\BigOSI[2]%
  {\mathcal{O}}{(}{)}{}%
  {\SI{#1}{#2}}

\begin{document}

\preprint{APS/123-QED}

\title{Noise in the direction of motion determines the spatial distribution and proliferation of migrating cell collectives}
\author{Jonathan E. Dawson} 
\author{Abdul N. Malmi-Kakkada}
\email[Corresponding author:]{amalmikakkada@augusta.edu}
\affiliation{Department of Physics and Biophysics, Augusta University, Augusta, GA 30912, USA}
\date{\today}

\begin{abstract} 
A variety of living and non-living systems exhibit collective motion.
From swarm robotics to bacterial swarms, and tissue wound healing to human crowds, examples of collective motion are highly diverse but all of them share the common necessary ingredient of moving and interacting agents.   
While collective motion has been extensively studied in non-proliferating systems, how the proliferation of constituent agents affects their collective behavior is not well understood.  Here, we focus on growing  active agents as a model for cells and study how the interplay between noise in their direction 
of movement and proliferation determines the overall spatial pattern of collective motion.  
In this agent-based model, motile cells possess the ability to adhere to each other through cell-cell adhesion, grow in size and divide. 
Cell-cell interactions influence not only the direction of cell movement but also cell growth through a force-dependent mechanical feedback process. 
We show that noise in the direction of a cell's motion has striking effects on the emergent spatial distribution of cell collectives and  proliferation. While higher noise strength leads to a random spatial distribution of cells, we also observe increased cell proliferation. On the other hand, low noise strength leads to a ring-like spatial distribution of cell collectives together with lower proliferation.  
Our findings provide insight into how noise in the direction of cell motion determines the local spatial organization of cells with consequent mechanical feedback on cell division impacting cell proliferation due to the formation of cell clusters.   
\end{abstract}

\maketitle

\section{Introduction}
The importance of the coordination between 
cell division and cell migration is recognized in multiple physiological processes, such as tissue regeneration, inflammation, as well as in pathological conditions, such as cancer metastasis~\cite{wu2012gradient,yang2017promoting}. 
Because cell migratory and proliferation patterns determine how cells organize spatially over time, understanding the underlying biophysical mechanisms is crucial for our ability to direct spatial organization of cells in a customizable manner. 
This has important implications 
for understanding tissue regeneration and cancer invasion~\cite{yang2017promoting, Shim2021}. 

With the emergence of multiplexed tissue imaging modalities that allow for quantification of cell proliferation at single-cell resolution ~\cite{giesen2014highly,lin2018highly}, it is now possible to determine how cell-cell interactions influence cell proliferation~\cite{gaglia2022temporal} from spatial map of single cells, together with higher-order relationships in space. 
In a cell collective, spatial constraints due to crowding limits the space available to a cell due to the presence of neighboring cells and thus impose constraints on cell proliferation~\cite{shraiman2005mechanical, streichan2014spatial, malmi2022adhesion}. 
Similarly, collective cell migration, a foundational collective behavior  in living systems, involves 
both the interaction of a cell with its environment as well as its neighbors \cite{Ladoux2017, Angelini2010, Shim2021, Pascalis2017, Mehes2014}. 
Fluctuations in the direction of a cell's motion 
affects the spatial coordination of cells in a tissue~\cite{Shim2021,Kuriyama2014}. 
Despite the importance of cell-cell interactions, the relation between cell migration driven spatial organization and how it impacts cell proliferation due to physical constraints remains unclear.
Given that cells are active particles that transduce stored energy into mechanical motion, an interesting question that arises is how the coordination between cell migration and proliferation influences the spatial organization of cell collectives.
While cell growth, cell division, and cell migration are highly complex processes, involving a large network of intracellular signaling pathways \cite{Huang2004}, here we focus on the  
biophysical intercellular interactions that are known to play a key role in cell collective migration and proliferation \cite{Melani2008, malmi2018cell, sinha2020spatially, sinha2021inter, Dawson2022b}.   

Mathematical and computational models of cell behaviors have contributed to a quantitative understanding of collective cell migratory behaviors and its underlying mechanisms~\cite{Mehes2014, Yang2014, Woods2014, Battersby2015, Gregoire2003, Buttenschon2020,huebner2021mechanical}. 
Pioneering work by Vicsek and co-workers showed that the collective dynamics of self-driven, or active particles emerge from a form of inter-particle coupling: a simple rule that an individual constituents' direction of motion is aligned  with the average direction of motion of its neighbors ~\cite{vicsek1995novel}. 
Both the number density of agents and noise in the direction of their movement are key parameters that regulate spatial patterns of collective motion. 
Distinct from earlier studies, we focus on studying the coupling between noise in the directionality of cell migration and cell division. 
The effect of cell division and cell death on collective cell movement has been studied in mean-field dynamical theoretical models \cite{Binny2015, Mishra2022} with recent experiments showing that cell growth and division can influence cell migratory behavior  \cite{Heinrich2020}. 
Our recent work 
in the context of freely expanding three-dimensional (3D) cell collectives~\cite{Malmi-Kakkada2018,sinha2020spatially} showed that the inter-cellular forces give rise to heterogenous cell motility patterns between the boundary and the interior of the cell collective. 
In addition to cell-cell mechanical interactions, we anticipate that the noise in the cell movement direction may generate complex spatial distribution patterns with novel 
implications on how cells divide.  

To elucidate the role of noise on self-organization and proliferation in a migrating cell  collective, we study a system of self-propelled particles with the capacity to proliferate, and whose motion is governed by local alignment rules. 
Each cell can grow in size and divide upon reaching a critical size. 
Cells in direct contact through cell-cell adhesion exert a force, which when exceeds a threshold inhibits cell growth and prevents cell division. 
Such mechanical feedback on cell proliferation is in agreement with recently reported experimental observations ~\cite{di2022pressure}. 
Cell division events in this model scramble the velocity orientation of dividing cells. 
By combining mechanical and alignment cell-cell interactions with cell division events, our model is highly relevant to biological systems, such as cells, which possess an inherent capability to proliferate and migrate. 
Our work provides insight into the fundamental features of expanding active matter. 
Notably, we discover that noise in the direction of a cell's motion not only influences the spatial structure of cell collectives but also determines the ability of cells to proliferate. 

\section{Model description and simulation details}
Here we introduce the computational model we implemented to study the growth and migration of cell collectives in two-dimensions (2D).   
The off-lattice agent-based model and the simulation scheme is adapted from our previous work on three-dimensional 
tumor growth \cite{malmi2018cell, sinha2020spatially, sinha2021inter,  malmi2022adhesion, zills2023enhanced}.  
Such off-lattice simulations are widely used to recapitulate experimentally observed features of individual cell dynamics within cell collectives \cite{drasdo2005single, schaller2005multicellular}. 
\begin{figure*}[ht]
     \centering
\includegraphics[width=0.8\textwidth]{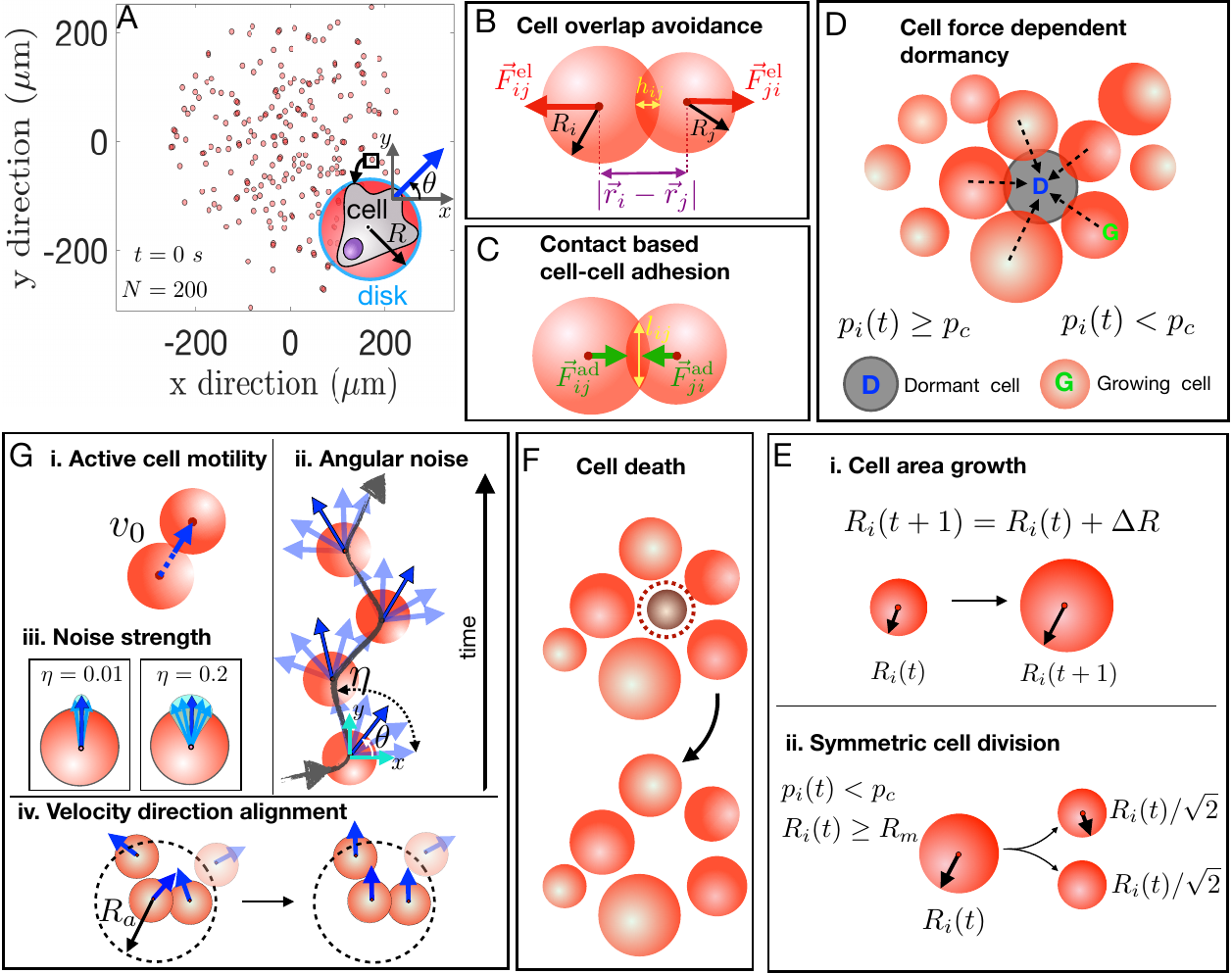}
     \caption{\footnotesize \textbf{Self-propelled particle model for a two-dimensional (2D) proliferating cell collective.}  
     Cells are represented as motile 2D circular disks that move with an active speed of $v_0$. Individual cells can grow in size and divide over time. 
     \textbf{(A)} Initial snapshot of the simulation is shown where each dot represents a cell (see inset). Cell motility leads to the spatial spread of the collective, which is quantified by tracking individual cell positions. 
     \textbf{(B)}  When two cells overlap, each experience an elastic force ${\bf F}^{el}_{ij}$, where $i$ and $j$ are the cell indices. The magnitude of the elastic force is  proportional to the degree of overlap $h_{ij}$.
     \textbf{(C)} Cell adhesion in the model is mediated by receptors on the cell membrane. In this simplistic model, where the receptors and ligands are assumed to be evenly distributed on the cell surface, the magnitude of the adhesive force 
     is assumed to scale as a function of their contact length $l_{ij}$. \textbf{(D)} If the force per unit length $p_i$ that the $i$th cell experiences (due to neighbor cell contacts) exceeds a critical threshold $p_c$, i.e., $p_i(t)\geq p_c$ then that cell enters the dormant state (D). 
     \textbf{(E)} (i-ii) If $p_i(t)< p_c$, the cells grow (G) until they reach the mitotic radius $R_m$. At that stage, the cell undergoes area conserved symmetric cell division, giving rise to two new cells with the same radii. A cell that is dormant at a given time can transit from that state at subsequent times. 
     \textbf{(F)} The cells can undergo apoptosis at a rate $k_d$. Upon apoptosis, the cell is removed from the cell collective. 
     \textbf{(G)} Together with proliferation, the cells are endowed with the capability to self-propel. 
     (i) Motility of each cell in the collective is characterized by a constant active speed $v_0$ along the direction, or orientation, of motion $\theta$. 
     (ii-iv) The net direction of cell movement depends on the angular noise (light blue arrows). 
     The strength of the angular noise $\eta$ determines how much a cell deviates from its current direction of motion in the next instant. 
     Additionally, the direction of motion of a cell depends on all of its nearest neighbors over a spatial range of length $R_a$ together with the net intercellular interaction force (due to repulsion and adhesion). 
}
     \label{fig:model_intro}
\end{figure*}
Individual cells are modeled as soft disk-like motile particles of radius $R$, Fig.\ref{fig:model_intro}A, which grow stochastically in time $t$, and, upon reaching a critical size, undergo division into two daughter cells. 
In addition to its radius, $R_i(t)$, the state of each cell $i$ is characterized 
by its position $\mathbf{r}_i(t)$ and direction of motion $\theta_i(t)$, Fig.\ref{fig:model_intro}A (Inset). 
The dynamics of the proliferating and migrating cell collective is governed by the following three factors - (a) mechanical forces arising from two body interactions, (b) active processes due to cell growth, division, and death, and (c) active self-propulsion with directional noise together with neighbor interactions that align the direction of cell motion with its neighbors. 
The model implementation of these factors is explained in detail below. 

{\bf (a) Mechanical cell-cell interactions:} Individual cells interact with short-ranged forces, consisting of two terms: elastic force (repulsion) and adhesion (attraction). 
The elastic force, $F_{ij}^{el}$, between any pair of cells $i$ and $j$ of radii $R_i$ and $R_j$ 
discourages spatial overlap between cells (Fig.\ref{fig:model_intro}B) and is given by~\cite{schaller2005multicellular,malmi2018cell},
\begin{equation}
F_{ij}^{el}=\frac{h_{ij}^{3/2}}{\frac{3}{4}(\frac{1-\nu_i^2}{E_i}+\frac{1-\nu_j^2}{E_j})\sqrt{\frac{1}{R_i(t)}+\frac{1}{R_j(t)}}},
\label{eq:f_repulsion}
\end{equation}
where $\nu_i$ and $E_i$ are the Poisson ratio and elastic modulus of the $i^{th}$ particle. $h_{ij}$ defined as $\mathrm{max}[0,R_i + R_j - |\vec{r}_i - \vec{r}_j|]$ is the virtual overlap distance between the 
two cells~\cite{malmi2018cell}. 
Biological cells adhere to their immediate physical neighbors through 
cell adhesion molecules, Fig.\ref{fig:model_intro}(C).
The adhesive force, $F_{ij}^{ad}$, between a pair of interacting cells  depends on the contact length between two cells, $l_{ij}$ (see Supplemental Information SI-I for the analytical calculation of $l_{ij}$), and is given by~\cite{schaller2005multicellular,malmi2018cell,malmi2022adhesion}, 
\begin{equation}
F_{ij}^{ad}=f^{ad}l_{ij}\frac{1}{2}(c_{i}^{rec}c_{j}^{lig} + c_{j}^{rec}c_{i}^{lig})
\label{eq:f_adhesion}
\end{equation}
where,  
and $c_{i}^{rec}$ ($c_{i}^{lig}$) is the receptor (ligand) concentration 
(assumed to be normalized with respect to the maximum receptor or ligand concentration so that  
$0 \leq c_{i}^{rec},  c_{i}^{lig} \leq 1$). 
The coupling constant $f^{ad}$ allows us to 
rescale the adhesion force to account for the variabilities in the maximum densities of the receptor and ligand concentrations. 
\begin{table}
\begin{center}
\begin{tabular}{ |p{4.5cm}||p{3cm}|p{1cm}|  }
 \hline
 \bf{Parameters} & \bf{Values}  \\
 \hline
 Timestep ($\Delta t$)& 5$\mathrm{s}$  \\
  \hline
Active cell speed ($v_{o}$) &  0.1 $\mathrm{\mu m}/s$ \\
 \hline
Critical Radius for Division ($R_{m}$) &  5 $\mathrm{\mu m}$ \\
 \hline
Environment Viscosity ($\mu$) & 0.005 $\mathrm{kg/ (\mu m~s)}$  \\
 \hline
Adhesive Friction Coefficient ($\zeta^{max}$) & $10^{-4} \mathrm{kg/ (\mu m~s)}$ \\
\hline
 Benchmark Cell Cycle Time ($\tau$)  & 1000 $\mathrm{s}$\\
 \hline
 Adhesive Coefficient ($f^{ad})$&  $10^{-4} \mathrm{\mu N/\mu m^{2}}$\\
 \hline
Mean Cell Elastic Modulus ($E_{i}) $ & $10^{-3} \mathrm{MPa}$ \\
 \hline
Mean Cell Poisson Ratio ($\nu_{i}$) & 0.5 \\
 \hline
 Death Rate ($k_d$) & $10^{-20} \mathrm{s^{-1}}$ \\
 \hline
Mean Receptor Concentration ($c^{rec}$) & 0.9 (Normalized) \\
\hline
Mean Ligand Concentration ($c^{lig}$) & 0.9 (Normalized)\\
\hline
Threshold force ($p_c$) & $10^{-4} \mathrm{MPa}$ \\
\hline
Noise strengths ($\eta$) & $0.01-0.2$ \\
\hline
\end{tabular}
\label{table_1}
\caption{The parameters used in the simulation.}
\end{center}
\end{table}

Both the elastic and the adhesive forces act along the unit vector $\textbf{n}_{ij}$, pointing from the center of cell $j$ to the center of cell $i$. 
The net force (${\bf F}_i$) on the $i^{th}$ cell is the vectorial sum of the elastic and adhesive forces that the neighboring cells exert on it,
\begin{equation}
{\bf F}_i=\sum_{j\in NN(i)} {\bf f}_{ij}=\sum_{j\in NN(i)}(F_{ij}^{\textrm{el}}-F_{ij}^{\textrm{ad}})\textbf{n}_{ij}
\label{eq:netforce}
\end{equation}
here, $j$ is summed over the number of nearest neighbors $NN(i)$ of cell $i$. 
The nearest neighbors of cell $i$ are all the cells that satisfy the criterion $h_{ij}>0$. 
The net force due to finite area exclusion (elastic term) and cell-cell adhesion is dampened by an effective friction contribution which comes from (i) the interaction of a cell with the extracellular matrix (ECM), and (ii) cell-cell adhesion. 
The friction that a cell $i$ experiences is a time ($t$) dependent quantity given by, 
\begin{equation}
\gamma_i(t) = \gamma_i^{ECM}(t) + \gamma_i^{\textrm{ad}}(t)\ \ . 
\label{eq:friction_1}
\end{equation}
The cell-ECM friction coefficient is assumed to be given by the modified Stokes relation,
\begin{equation}
\gamma_i^{ECM}(t)= \mu R_i(t),
\label{eq:friction_2}
\end{equation}
where, $\mu$ is the viscosity due to the ECM. 
We consider additional damping of cell 
movement due to adhesive forces given by, 
\begin{eqnarray}
\gamma_i^{\textrm{ad}}=&& \zeta^{\textrm{max}}\sum_{j \in NN(i)}\bigg(\frac{l_{ij}}{2}(1+\frac{\mathbf{F}_i\cdot \mathbf{n}_{ij}}{|\mathbf{F}_i|})\times \\ \nonumber &&\frac{1}{2}(c_{i}^{rec}c_{j}^{lig} + c_{j}^{rec}c_{i}^{lig})\bigg)
\label{eq:damping_adhesion}
\end{eqnarray}
where, $\zeta^{max}$ is the adhesive friction coefficient and $\mathbf{F}_i$ is as defined in Eq.(\ref{eq:netforce}). 
Note that the added friction coefficient $\gamma_{i}^{\textrm{ad}}$ is proportional to the cell-cell contact length $l_{ij}$, implying that the damping of cell movement due to this friction term is proportional to the number of cells that cell $i$ is in contact with at time $t$. 

{\bf (b) Cell proliferation:} In our model, the cell number grows due to the imbalance between cell division and apoptosis. 
At any point in time, cells are either in the growth (G) phase, i.e, the phase in which the cell area increases over time, or, in the dormant (D) phase, i.e., the phase in which cell area growth is arrested, Fig.\ref{fig:model_intro}D. 
Whether a cell continues in the growth phase or enters the dormant phase is determined by the total force per unit length, due to the neighboring cells, acting on a cell at any given time point. 
The total external force per unit length, $p_i$, that a cell experiences is calculated using, 
\begin{equation}
    p_i(t)=\sum_{j\in NN(i)}\frac{|{\bf f}_{ij}\cdot {\bf n}_{ij}|}{l_{ij}}. 
    \label{eq:pressure}
\end{equation}

If $p_i(t)$ on a cell $i$ at any given time $t$ is smaller than a threshold value, $p_c$, the cell grows in size, Fig.\ref{fig:model_intro}E-i. 
However, if $p_i(t) > p_c$, the cell enters dormancy, Fig.\ref{fig:model_intro}D. 
Hence, depending on the ratio of $\frac{p_i (t)}{p_c}$, cells can switch between the two states of dormancy and area growth. 
A cell grows in size by increasing its radius in a stochastic manner sampled from a Gaussian distribution with the mean rate $\frac{dR_i}{dt}= (2\pi R_i)^{-1}g_a$, where $g_a$ is the cell area growth rate given by, 
\begin{equation}
    g_a= \frac{\pi R_m^2}{2\tau}.\
    \label{eq:growthrate}
\end{equation}
Here, $\tau$ is the cell cycle time and $R_m$ is the mitotic radius at which a cell divides (see Table I). 
We assume that a cell divides into two daughter cells upon reaching  $R_m=5\mu$m, giving rise to two identical daughter cells, each with radii $R_d=\frac{R_m}{\sqrt{2}}$, ensuring area conservation, Fig.\ref{fig:model_intro}E-ii.  
Hence, a key time scale in the simulation is $\tau$ - the average time it takes for a cell to divide, set to be $\sim 0.27~$hours. This is much faster than the typical cell cycle times of eukaryotic cells but comparable to cell cycle times of bacteria~\cite{shamir2016snapshot}. 
As daughter cells are assigned completely random active velocity orientations, cell division events tend to scramble the orientational order of the cells.  
Death of a cell takes place in the simulation leading to a randomly selected cell being removed from the collective, Fig.\ref{fig:model_intro}F. 
The death rate is set to $k_d=10^{-20} s^{-1}$. Owing to $k_d << \frac{1}{\tau}$, we are simulating a rapidly growing system of cells.

{\bf (c) Neighbor velocity alignment and fluctuation in the direction of motion:} 
The cell position, $\mathbf{r}_i(t)$, is described through the coordinates $(x_i(t), y_i(t))$. Cell  
self-propulsion velocity is, $\boldsymbol{v}_i(t)=v_0\boldsymbol{s}_i(t)$, where, $v_0$ is the cell migration speed, and $\boldsymbol{s}_{i}(t)=(\cos \theta_i(t), \sin \theta_i(t))$ is the unit vector representing the direction of cell migration. The angle that the cell makes with the horizontal axis in the laboratory frame is $\theta$. 
Each cell in this model is endowed with motility that propels the cell in a given direction with a fixed speed $v_0$, Fig.\ref{fig:model_intro}G-i. 

The directional alignment, and thus the overall direction of a cell's motion, is hampered by an angular white noise uniformly distributed in range $\xi_i \in [-\frac{\pi}{2}, +\frac{\pi}{2}]$ with $\langle\xi_{i}^t\rangle=0$ and $\langle \xi_{i}^t \xi_{j}^{t'}\rangle \sim \delta_{ij}\delta_{tt'}$ and whose strength is given by $\eta$, Fig.\ref{fig:model_intro}G-ii. As the effective noise is given by $\eta \xi_i$, 
$\eta=0.2$ means random fluctuations occur in the entire range $[-\frac{\pi}{10}, +\frac{\pi}{10}]$, Fig.\ref{fig:model_intro}G-iii, whereas, $\eta=0.01$ results in random fluctuations in the range $[-\frac{\pi}{200}, +\frac{\pi}{200}]$, Fig.\ref{fig:model_intro}G-iii. 
The noise term represents fluctuations in the direction of a cell's motion. In biological systems, such as cells, there are many sources of such noise in the direction or orientation of cell movement.  
Stochasticity intrinsic to cellular movement, such as due to limitations in cellular sensing or active shape remodeling during cell migration \cite{Chen2014, Caballero2014} are some examples. 

In addition to the forces due to nearest neighbor mechanical interactions, as described in (a), each cell interacts with its neighbors in a manner that aligns its own velocity with that of its neighbors, Fig.\ref{fig:model_intro}G-iv. 
The nearest neighbors which contribute to the velocity re-alignment of cell $i$ are all those cells in the collective that satisfy the necessary condition $|\boldsymbol{r}_i({t})-\boldsymbol{r}_j({t})|< R_a$, where, $|{\bf ...}|$ is the vector magnitude, Fig.\ref{fig:model_intro}G. 
We set $R_a=10 \mu m$ which limits 
velocity re-alignment to occur with neighbors that are directly in contact with a given cell. 
We then obtain the average orientation of the velocities of all the cells that satisfy the nearest neighbor criteria and assign that to the velocity orientation of cell $i$. 
The cell velocity re-alignment with its neighbors influences its direction of motility, such that cells in a cluster tend to move in the same direction, Fig.\ref{fig:model_intro}G-iv. 
Contact-based modulation of cell velocity is known to play a role in the collective migration of electrically stimulated cells \cite{Dawson2022b}. 

The complex dynamics of each cell in the collective involves active motility, area growth, division, and death. In the low Reynolds number limit, the equation of motion is fully described by the following update rules:
\begin{eqnarray}
r^x_i(t+\Delta t)&=&r^x_i(t)+v_0 \textrm{cos}(\theta_{i}(t))\Delta t+\frac{ F^x_i(t)}{\gamma_i(t)}\Delta t \label{eq:drdt1} \\
r^y_i(t+\Delta t)&=&r^y_i(t)+v_0 \textrm{sin}(\theta_{i}(t))\Delta t+\frac{F^y_i(t)}{\gamma_i(t)}\Delta t \label{eq:drdt1b} \\
{\theta}_i(t+\Delta t)&=&  \textrm{arg}\big[\sum_{j \in |{\bf r}_i({t})-{\bf r}_j({t})|< R_a}{\bf s}_{j}(t)+ \sum_{j\in NN(i)}{\bf f}_{ij}\big] \nonumber \\
 && + \eta \xi_{i}(t)  \quad.
\label{eq:drdt2}
\end{eqnarray}

Eq.(\ref{eq:drdt1}-\ref{eq:drdt1b}) describes the evolution of the $x$ and $y$ coordinates of a cell 
$i$, governed by an active component that propels the cell with a speed $v_0$ in the direction ${\theta}_i(t)$ at time $t$ and the net force on the cell due to its contacting neighbors. 
We assume that the cell exerts a self-propulsion force which propels it with a constant effective active speed $v_o$. We note that an effective friction term is incorporated into the value of $v_o$. 
Eq.(\ref{eq:drdt2}) describes the orientation dynamics of a cell $i$, where $\theta_i(t+\Delta t)$ is the direction in which the cell moves in the next time step. 
The net contribution to the direction of a cell's motility comes from, (i) orientation re-alignment, the first term on the right-hand side of Eq.(\ref{eq:drdt2}) and, (ii) the interaction forces (discussed in (a)), second term on the right-hand side of Eq.(\ref{eq:drdt2}). 
As discussed in (c), the orientation re-alignment of a cell $i$'s velocity is only due to nearest neighbor cells whose center lies within a distance of $R_a$ (here $10 \mu m$) from the $i$th cell. 
$\textrm{arg}[\boldsymbol c]$ in the first term in Eq.(\ref{eq:drdt2}) refers to the angle associated with the vector ${\bf c}$, if this is expressed in polar coordinates, and the sum is taken over all cells $j$ within a distance of $R_a$ of cell $i$ (including cell $i$ itself). 
The net direction in which cell $i$ moves is given by the angle associated with the net vector, which is obtained by vector addition of the velocity vectors of all neighboring cells which lie within the interaction radius $R_a$ of cell $i$, and the net force ${\bf F}_i$ on the $i$th cell.

\begin{figure*}[t!]
     \centering
\includegraphics[width=0.8\textwidth]{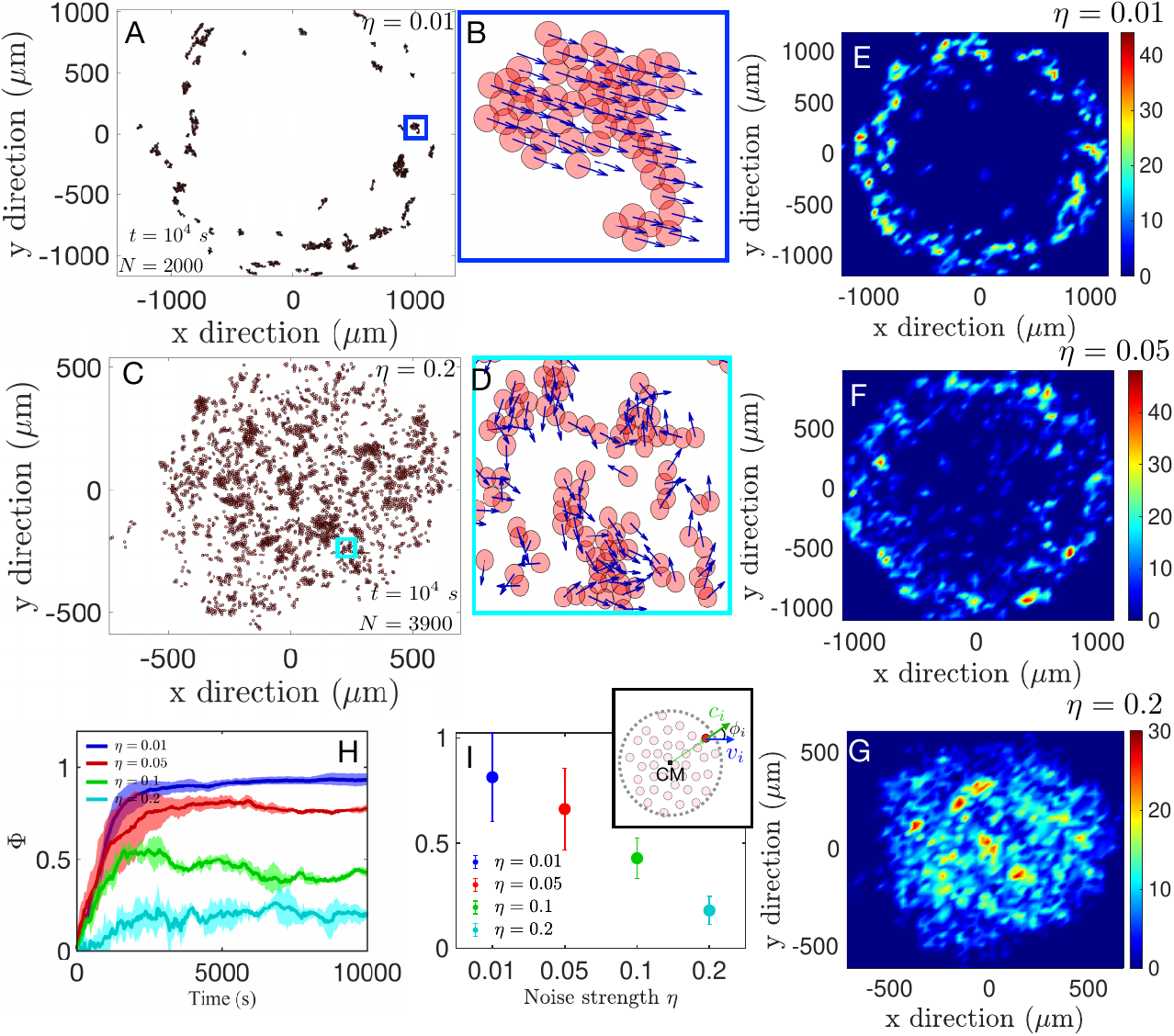}
     \caption{\footnotesize \textbf{Noise determines the spatial distribution patterns  and collective motion of growing cell collectives.}  
\textbf{(A-B)} Spatial distribution of cells at $t=10^4 s$ for low angular noise in the direction of motion, $\eta=0.01$. Starting from 200 cells, the cell collective grows to $\sim$2000 cells distributed amongst multiple smaller clusters. Multiple clusters of cells are observed which self-organize into a ring-like pattern, which expands over time as cells proliferate and migrate. The interior of this domain is mostly devoid of cells. Within each cluster (for example, see the blue box and magnified view in panel B), cell movement is tightly aligned. 
\textbf{(C-D)} Spatial distribution of cells at high angular noise in the direction motion $\eta=0.2$, at the final simulation timepoint $t=10^4 s$. Starting from 200 cells, there are $\sim3900$ cells at the final time point. Multiple cell clusters are visible, however, these   
are dispersed all over the spatial domain with the velocity vectors of individual cells appearing more or less to be randomly oriented (see the cyan box and D showing a magnified view of a selected spatial region). 
\textbf{(E-G)} Density plot of the spatial distribution of cells in the collective for (E) $\eta=0.01$, (F) $\eta=0.05$, and (G) $\eta=0.2$. 
The 2D space is divided into a $50\times50$ grid. Colormap shows the number of cells in each 2D spatial bin. 
The difference in the spatial distribution pattern of cells is clearly visible, as the noise strength is increased from (E) $\eta=0.01$ to (G) $\eta=0.2$. 
At low noise, cells are densely packed in a narrow ring-like region. The interior part of the domain in which the cells are spread is largely devoid of cells. At high noise strength, $\eta=0.2$, cells are densely packed in the central part of the circular domain over which they are spread. 
{\bf(H)} The degree of collective cell motion is quantified using the order parameter $\Phi$ which is obtained (see inset of (I)) from the cell velocity vector and cell position vector relative to the center of mass of the cell collective (see main text for the mathematical definition of the order parameter). 
$\Phi=0$ indicates highly disordered motion, in which cells form groups moving coherently in random directions. 
The temporal behavior of the order parameter shows an initial increase over time and gradually saturates at later times. 
 For high noise strength, $\eta=0.2$, the order parameter for the cell collective saturates at a value closer to 0. Whereas for low noise strength, $\eta=0.01$, $\Phi\sim1$. The thicker line denotes the mean value and the shaded area is the standard deviation here and henceforth. 
 {\bf(I)} Order parameter of cell collective at the final time point decreases as a function of the noise strength $\eta$. Inset visualizes the collective order parameter obtained from the relative outward orientation of cell motion $\phi_i$. 
}
     \label{fig:densitypolarplot}
\end{figure*}


{\bf Initial Conditions:} We initiated the simulations by generating 200 non-overlapping cells, randomly distributed in a circular region within a 2D spatial domain of size $250 \mu m \times 250 \mu m$. For all future time steps, we consider an open boundary condition. 
Each cell is assigned an initial orientation of the active velocity, randomly distributed in the domain $[0,2\pi]$. 
Fluctuations around the direction of a cell's motion is captured by a noise term, which is randomly distributed with uniform probability in the range $[\frac{-\pi}{2}, \frac{\pi}{2}]$. 
The strength of the fluctuations is denoted by $\eta$ (discussed in the previous section (c)).
In the present study, all the parameters are fixed except the noise 
strength of velocity orientation switching $\eta$, which we vary from $0.01$ to $0.2$. The simulated cell aggregate is evolved to $\sim 10\tau$ or 
about $10,000s$. Relevant parameters are shown in Table~I. A fixed timestep of $5$ s was used. We performed a numerical consistency check by 
ensuring our results are invariant for a smaller timestep of $2.5$s (see Supplemental Information SI-II). The 
particle coordinates were recorded and used to calculate the dynamical observables relevant to the present study.

\section{Noise in the cell motility direction controls the spatial distribution of the cell collective}
We first sought to understand how noise in the cell motility direction 
determines the spatial distribution 
of a growing cell collective. 
The cell spatial distribution that we obtain at $t= 10,000 s$ shows a strong dependence on the noise strength $\eta$, Fig. \ref{fig:densitypolarplot}A, C. 
For low noise strengths ($\eta=0.01$) cells are organized into multiple clusters that are spatially distributed in a roughly circular, ring-like pattern 
Fig.\ref{fig:densitypolarplot}A (see Supplemental Information SI-III for simulation movies). 
The cells cluster into small groups mostly along the edge of the ring-like domain. 
The domain interior is mostly devoid of cells, Fig.\ref{fig:densitypolarplot}A. 
By focusing on a single cluster (blue box in Fig.\ref{fig:densitypolarplot}A), we observe that the constituent cells display highly coordinated motion, wherein each cell moves in roughly the same direction pointing radially outward, as seen from the blue arrows in Fig.\ref{fig:densitypolarplot}A(inset), B. 
At higher noise strength of $\eta=0.2$ the cell spatial distribution 
changes from the ring-like structure to a diffuse morphology, characterized by randomized spatial distribution of cells, Fig.\ref{fig:densitypolarplot}C (see Supplemental Information SI-III for simulation movie). 
The cells organize into a large number of clusters of varying sizes scattered throughout the entire spatial domain occupied by the cells, Fig.\ref{fig:densitypolarplot}C. 
Individual cells within each cluster appear to move in a less coordinated manner, as compared to the case of low noise strength, Fig.\ref{fig:densitypolarplot}C,D.  
To better visualize the differences in the cell spatial distribution and 
the cluster sizes at varying $\eta$, we represented the cell positional information  using a density plot. 
The entire spatial domain, in both $x$ and $y$ direction, is divided into $50\times 50$ bins of equal area. 
The total number of cells within each bin is color-coded, with dark blue representing low number of cells and dark red representing the highest number of cells. 
To generate the cell number density heat map, we combined 3 separate simulation results  for each value of the noise strength, $\eta=0.01, 0.05, 0.2$, Fig.\ref{fig:densitypolarplot}E-G.  
The density plots show clearly the strong influence of the noise strength on the cell spatial distribution.  
For low noise strength, $\eta=0.01$, the whole collective is spatially organized into a thin circular ring-like structure, with patches of high cell density visible at the border. The interior of the domain is characterized by low cell number density, Fig. \ref{fig:densitypolarplot}E.  
Cells organize themselves into coherently moving clusters with some of the larger clusters containing about 40-50 cells as seen in Fig. \ref{fig:densitypolarplot}E. 
At higher noise strengths of $\eta=0.05$ and $0.2$, high cell density patches shift from being confined to the border of the ring-like pattern to its interior. 
The number of cells within the high cell density patches decreases in a noise strength dependent manner. While 40-50 cells make up the high-density patches for $\eta=0.01$, $\sim 30$ cells are visible for $\eta=0.2$.  
The cell spatial distribution we observe is not a transient feature of the model.  
Long-time simulations (upto $t=25,000\ s$), for $\eta=0.01$ and $\eta=0.2$ (see Supplemental Information SI-IV) confirm that the cell spatial distribution 
is preserved even after very long times. 
We, therefore, conclude that the noise-dependent pattern of cell collective behavior is a robust feature of expanding cell collectives. 

The velocity vector alignment of individual cells within a cluster, seen in 
Fig.\ref{fig:densitypolarplot}B and D, are indicative of collective behavior 
seen in non-proliferating 
self-propelled particles \cite{vicsek1995novel}. 
To better understand the collective motion of individual cells, we measured the order in the motion of the entire cell collective (Fig.\ref{fig:densitypolarplot}H). 
We calculate the order parameter on the basis of position-dependent polarization of the cell velocity  
by defining a vector pointing 
from the center of mass of the cell collective to the individual cell position ${\bf c}_i = {\bf r}_i - {\bf R}_{CM}$, where ${\bf R}_{CM}(t)=(1/N)\sum_i {\bf r}_i$ is the center of mass of the whole collective at time $t$. 
${ \bf c}_i$ is directed outwards from the center of mass of the entire cell collective to the cell's position. 
The angle $\phi_i$ between a cell's velocity vector, ${\bf v}_i$, and its position vector with respect to the center of mass of the cell collective, ${\bf c}_i$, can be calculated from 
$\cos(\phi)_i = {\bf c}_i \cdot {\bf v}_i/(|{\bf c}_i||{\bf v}_i|)$ (see Fig.\ref{fig:densitypolarplot}I Inset). 
The orientation order parameter for the whole cell collective at any given time $t$ is defined as, 
\begin{equation}
\Phi(t) = \frac{1}{N(t)}\sum_i \cos(\phi(t))_i
\label{eq:orderparam}
\end{equation}
where, $N$ is the total number of cells at time $t$. 
$\Phi$ can vary between 1 and 0 with 
$\Phi=1$ implying that the velocity orientation ${\bf v}_i$ of each cell in the whole cell collective is aligned with respect to the position vector ${ \bf c}_i$. 
The time-dependent behavior of $\Phi(t)$ shows an initial almost linear increase over time which then saturates at a constant value at later times, Fig.\ref{fig:densitypolarplot}H. 
For very low noise strength of $\eta=0.01$, the order parameter saturates at $\sim1$, indicating a highly ordered outward cell motion.  
This is consistent with our observation of 
highly coherent and ordered cell movement such that cell velocity orientation ${\bf s}_i$ is aligned with the vector pointing outward towards the periphery of the cell collective, ${\bf c}_i$. With increasing noise strength, the value of the order parameter progressively gets lower, indicating an increasingly disordered velocity direction.  
The orientational order parameter at the final time point is shown in Fig.\ref{fig:densitypolarplot}I, clearly decreasing with higher noise strengths. 
Our result, showing the dependence of the order parameter on the noise strength, 
also delineates why we obtain markedly distinct spatial distribution of cell collectives. While cells move consistently outwards at low noise 
strengths leading to the emergence of a ring-like pattern, higher noise strengths result in randomized cell movement orientations that lead to a more diffuse spatial distribution of cells. 
In general, our results map out the emergent spatial distribution of proliferating  cell collectives. 


\section{Noise in the cell motility direction determines proliferation and the spread of cell collective}
Having observed angular noise-dependent differences in the spatial 
distribution and the orientational order of cell collectives, we next ventured to ask how the noise influences cell division and the growth of the cell collective. 
As spatial constraints can regulate cell cycle progression during tissue expansion~\cite{streichan2014spatial, di2022pressure, malmi2022adhesion}, we anticipate that noise-induced differences 
in the cell spatial distribution will have an impact on the ability of cells to divide. Particularly, given that we incorporate mechanical feedback on cell division through the force term, noise-induced differences in local cell spatial arrangements could determine the ability of cells to divide. 
\begin{figure}[htb]
    \centering
\includegraphics[width=0.48\textwidth]{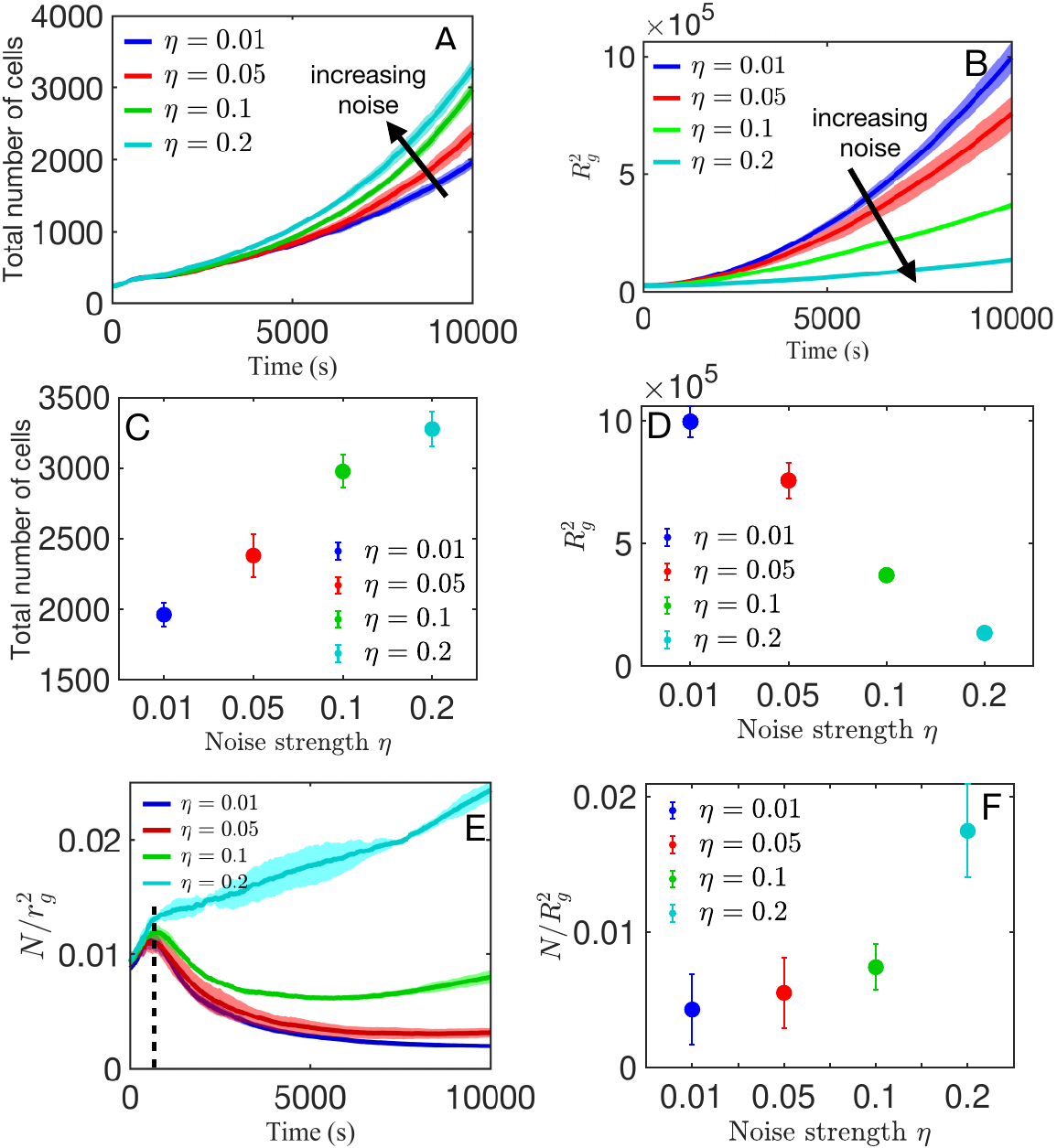}
 \caption{\textbf{Noise determines cell proliferation, spatial spread and packing of growing cell collectives.} {\bf (A)} Temporal behavior of the total number of cells. Although the cell number steadily increases for each value of the noise strength, the increase is more pronounced at  higher noise strengths. 
 {\bf(B)} Temporal behavior of the total spatial spread of cell collective quantified using the radius of gyration squared $R_g^2$ ($\mu m^2$). The spatial spread showed a steady increase for each value of noise strength, with the spatial spread enhanced at lower noise strengths.
 {\bf(C)} Total number of cells at the final time point $t=10,000 s$ as function of the noise strength $\eta$. The growth of the cell collective is enhanced at higher noise strengths. 
 {\bf(D)} Total spread of cell collective at the final time point $t=10,000 s$ as function of the noise strength $\eta$. The spatial spread of the cell collective is inversely proportional to the noise strength. 
 {\bf (E)} We quantify the dynamics of cell spatial packing by evaluating the temporal behavior of the cell number density. At the outset, initially, the cell number density shows a sharp rise over a short duration of time. Thereafter, the cell number density decays over time for noise strengths of $\eta=0.01,0.05$ and $0.1$. For $\eta=0.2$, the cell number density increases at late times. This increase is highly pronounced for $\eta=0.2$
 and continuously increases over time. 
 {\bf(F)} Cell number density at $t=10,000 s$ as a function of the noise strength $\eta$. On average, this indicates that cells are more tightly packed spatially with increasing noise strength. 
}
\label{fig:totnumberdensity}
\end{figure}
To understand how noise in the cell velocity orientation affects the  
proliferation of the cell collective, we looked at the temporal behavior of the total cell number and total spread area of the cell collective, for four different values of the noise strengths $\eta=0.01, 0.05, 0.1, 0.2$. 
We quantified the spatial spread of migrating cell collective by calculating the radius of gyration squared, 
\begin{equation}
 R_g^2(t)=\left\langle  \frac{1}{N} \Sigma_{i=1}^{N} [{\bf r}_{i}(t)-{\bf R}_{CM}(t)]^2 \right\rangle.
\label{eq:radg_sqrd}
\end{equation}
The bracket $\langle ... \rangle$ denotes the ensemble average over 3 different simulation runs at each value of $\eta$. 
The average squared distance of all the cells from the center of mass is an indicator of the spatial spread or invasion of a cell collective in two dimensions. 
Small $R_g^2$ values indicate a smaller spatial spread of cells, with cells localized in close proximity to the center of mass. 
In contrast, higher values of $R_g^2$ denote a wider spatial spread due to cells that are located farther away from the center of mass. 

Both the total number of cells, $N$, and the total spatial spread of cells, $R_g^2$, steadily increase with time, Fig.~\ref{fig:totnumberdensity}A,B for a given value of noise strength. 
In Fig.~\ref{fig:totnumberdensity}C,D, we show the $N$ and $R_g^2$ at the final time point. 
Surprisingly, at late time points $N$ and $R_g^2$ show opposite trends as a function of the noise strength $\eta$, Fig.~\ref{fig:totnumberdensity}C,D. 
The total cell number increases as the noise strength increases (see Fig.~\ref{fig:totnumberdensity}C), implying that stronger fluctuations in the direction of cell movement promote cell proliferation. At 
$t=10,000\mathrm{s}$, there are $\sim3400$ cells for $\eta=0.2$, while, $N\sim1900$ at the lower noise strength ($\eta=0.01$), which is significantly lower compared to the case of $\eta=0.2$, Fig.~\ref{fig:totnumberdensity}C. 
In contrast to the total number of cells, the total spatial spread of the cell collective showed an inverse dependence on the noise strength $\eta$. The spatial spread of the cell collective increases faster over time at lower noise strengths. 
$R_g^2$ is an order of magnitude smaller at $\eta=0.2$ as compared to the lower noise strength of $\eta=0.01$, suggesting that as the noise strength increases the cell collective exhibit a more compact spatial distribution (see Fig.~\ref{fig:totnumberdensity}D). 

The global quantities $N$ and $R_g^2$ describe the time-dependent behavior of the whole cell collective and how it is influenced by noise in the cell
motion direction. Taken together with the analysis presented in the preceding section, our results show that increasing the noise strength disrupts cell-cell 
velocity alignment, as reflected in the lower order parameter, but at the same time promotes cell proliferation, as reflected in the higher number of cells. On the other hand, lower noise strength facilitates cell-cell velocity alignment and 
suppresses cell proliferation.   

As collective behavior depends strongly on the number density of actively migrating agents~\cite{vicsek1995novel},  
we next sought to understand how cell number density is affected by noise in the direction of cell motility. 
Given that $N$ is not fixed and that we impose an open boundary condition, number density is neither fixed nor clearly defined, as in the case of Vicsek model, but evolves over time. 
Nevertheless, we can estimate the cell number density or the overall spatial packing of the cells using $\rho(t)=N(t)/R_g(t)^2$,  where $\rho$ is the cell density. 
Due to the combined effect of cell proliferation and cell motility, both the total number of cells $N(t)$ and the spatial spread $R_g^2$, evolve over time. 
Consequently, cell number density exhibits a highly dynamic time-dependent behavior.   
$\rho(t)$ initially increases sharply for each value of noise strength, $\eta=0.01, 0.05, 0.1$ and $0.2$, as shown in the time regime before the dashed line in Fig.~\ref{fig:totnumberdensity}E. 
Following the initial rise, the temporal profile of the cell number density for noise strengths $\eta=0.01, 0.05, 0.1$ is markedly different from that for $\eta=0.2$, Fig.~\ref{fig:totnumberdensity}E.   
For $\eta=0.01, 0.05, 0.1$, the cell number density decreases over time after the initial transient increase. 
Whereas for $\eta=0.2$, the cell density continues to increase with time, although at a lower rate. 
At longer times, cell number density is comparatively low for weaker noise strengths. 


By singling out the cell number density at the final time point and plotting it as a function of the noise strength, we show that the final cell density rapidly increases with the noise strength Fig.~\ref{fig:totnumberdensity}F. 
This dependence is rather surprising given our earlier results for the total number of cells as a function of noise strength. We expect higher proliferation to correspond to lower density, due to the role of cell contact force-dependent feedback on proliferation ($p_i (t)$) in our model. 
When cells are tightly packed in space, we expect the compressing forces on cells from their neighbors to be higher~\cite{malmi2018cell,malmi2022adhesion}. This would hamper cell area growth, eventually leading to lower cell division events due to the force-dependent mechanical feedback term $p_c$. 
Contrary to our expectations, high noise strength leads to a higher cell density and the cell collective has yet more number of cells (see Fig.~\ref{fig:totnumberdensity}A,C). To investigate this further, we turn to a more detailed quantification of the cell spatial arrangement on the basis of clustering analysis. 

\section{Noise increases the number of isolated cells and facilitates enhanced proliferation}

\begin{figure*}[t!]
    \centering
\includegraphics[width=0.95 \textwidth]{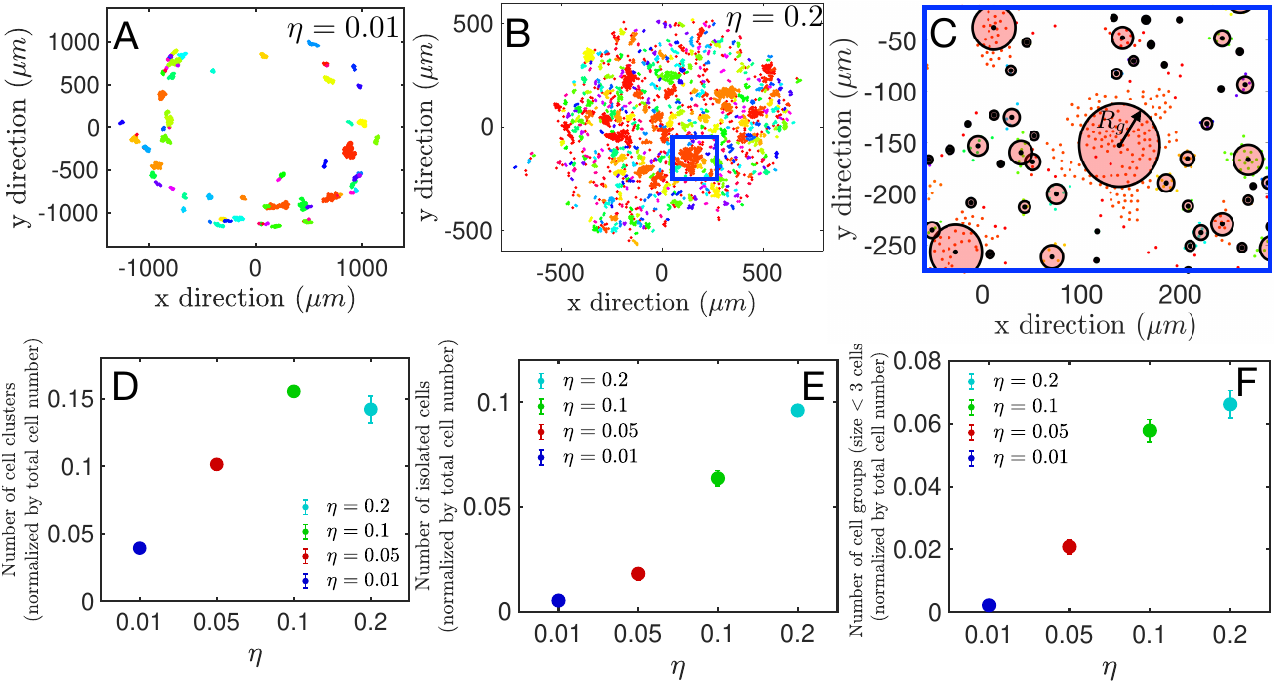}
\vspace{-3mm}
 \caption{ \textbf{Noise determines cluster partitioning of growing cell collectives.} Clustering analysis of cells using density-based spatial clustering of applications with noise (DBSCAN) organizes the whole cell population into distinct groups. 
Cells are either assigned to a cluster or labeled as an isolated cell. 
{\bf (A)} Cell clusters identified by DBSCAN for low noise strength $\eta=0.01$. Each cluster is given a unique color.
{\bf (B)} Cell clusters identified by DBSCAN for high noise strength $\eta=0.2$ with each cluster assigned a unique color.  
{\bf (C)} Magnified view of cell clusters with over-imposed circles whose radii reflect  cluster sizes. Cluster radius is quantified from the radius of the gyration, which is calculated by using the coordinates of the cluster center and the position of each cell in the cluster. 
{\bf (D)} Total number of cell clusters increase with noise strength $\eta$ (at the final simulation timepoint). 
{\bf (E-F)} Both (E) the number of single/isolated cells, i.e., cells that are not part of any cluster, as well as (F) the number of clusters with fewer than 3 cells increase with the noise strength $\eta$. 
}
\label{fig:dbscan}
\end{figure*}
To understand this rather counter-intuitive result of higher cell proliferation at higher cell number density, we used a spatial clustering algorithm DBSCAN (density-based spatial clustering of applications with noise) \cite{Ester1996} to map out the structure of cell clusters within the collective.  
The idea behind performing cluster analysis is that feedback due to the contact force from overlapping cells inhibits cell growth and hamper cell division. As such, single cells and cells with very few overlapping neighbors will be characterized by the highest proliferative capability. 
On the other hand, we expect fewer cell division events when cells are part of a cluster with larger number of overlapping cells. 
Therefore, we anticipate that the size of the cell clusters (i.e. the number of cells in a cluster) might hold the key to understanding why cells in a collective with higher global cell number density proliferate at a higher rate. 

DBSCAN is a powerful tool for class identification of clusters in large spatial databases with noise. 
For cluster identification and classification, DBSCAN requires two input parameters, namely, the maximum cell-cell distance $\epsilon [\mu\textrm{m}]$ to be considered as a cell's neighbor, and the minimum number of neighboring cells, $\textrm{n}_{\textrm{min}}$, that  qualify as a cluster. 
The DBSCAN algorithm initially labels each cell which has at least $\textrm{n}_{\textrm{min}}$ number of cells within a distance of $\epsilon [\mu\textrm{m}]$ from its center as a core cell. 
Any cell that has fewer than $\textrm{n}_{\textrm{min}}$ number of cells within a distance of $\epsilon [\mu\textrm{m}]$ from its center is labeled as border cell. 
All those cells which have no other cell in their neighborhood within a distance of $\epsilon [\mu\textrm{m}]$ from their center are labeled as single cells. 
The algorithm then randomly picks a core cell and assigns it a cluster index. 
The cluster is expanded sequentially, by adding cells which are in the neighborhood and within the distance of $\epsilon [\mu\textrm{m}]$ of the randomly picked core cell. In an iterative manner, DBSCAN algorithm labels each cell as being part of one of the clusters, with each cluster assigned a unique cluster index. 

Since only overlapping cells exert growth inhibiting force on each other, we focused on identifying cell clusters of overlapping cells. Therefore, and since the typical cell radii in our model is $5\mu$m, we chose $\epsilon=9\mu$m, which means that cell-center-to-cell-center distance between any two cells within a cluster is  $9\mu$m or less. 
This value of $\epsilon$ ensures that only overlapping cells form a cluster.
In order to cover the full range of cluster sizes we also set  $\textrm{n}_{\textrm{min}}=2$.  
Using MATLAB's in-built function for DBSCAN \cite{MATLAB}, with the aforementioned values for the two input parameters ($\epsilon$ and $\textrm{n}_{\textrm{min}}$), we identified cell clusters from spatial coordinates of individual cells at the final simulatiom timepoint and for different noise strengths $\eta$, Figs.~\ref{fig:dbscan} A,B. Each individual cell cluster in Figs.~\ref{fig:dbscan}A,B is represented in a different color. 
DBSCAN is a robust clustering method, allowing for the quantification of additional features of individual cell clusters. 
Based on the cluster identity of each cell, we can quantify the center of mass and the radius of gyration of individual cell clusters, as shown using circles of different radii in Fig.~\ref{fig:dbscan}(C). 

Our analysis shows that the entire cell collective is spatially organized into cell clusters of different sizes i.e. cell clusters are composed of varying cell numbers. Since the total number of cells varies with the noise strength, in order to perform cluster number comparison across different values of noise strengths, we normalized the total cell cluster number at a given noise strength by the total number of cells at that noise strength. 
The number of cell clusters at the final timepoint increases with the noise strength $\eta$, Fig.\ref{fig:dbscan} D. 
The slight dip in the cell cluster number at the highest noise strength of $\eta=0.2$ is due to a lower total number of clusters at $\eta=0.2$ as compared to $\eta=0.1$, which indicates that clusters tend to disintegrate into isolated or single cells when the value of $\eta$ is increased from $0.1$ to $0.2$. 
To understand higher proliferation in cell collective with higher cell number density we turned our attention to isolated cells and cell clusters with less than 3 cells.   
We found that the total number of both isolated cells and cell clusters with fewer than 3 cells increases with the noise strength $\eta$, Fig.\ref{fig:dbscan} E-F. 
These results are robust with respect to the simulation time, see Supplemental Information SI-V for simulations run for much longer time $t=25,000 s$. 
A higher number of isolated cells implies that more cells can proliferate, without the inhibitory effect of mechanical feedback on cell growth due to cell contact-dependent forces. This scenario is more conducive to cell division, allowing the cell collective to freely grow and divide.  

Our DBSCAN-based cell cluster analysis reveals that even though the cell number density is comparatively higher at higher noise strengths, there are large numbers  of isolated cells and clusters with fewer cell numbers. This leads to enhanced proliferation of individual cells.  
In an expanding cell collective, cells form clusters as a result of either cell-cell adhesion and/or nearest neighbor velocity alignment. 
As the noise strength increases, the tendency for these clusters to disintegrate or breakup increases, due to rapid fluctuations in the direction of migration. The isolated or smaller size clusters then proliferate at a higher rate, thereby increasing the total cell number even 
though the overall number density of cells is higher at higher noise strengths. Hence, locally, due to the presence of more cells with fewer 
neighbors, cells are able to grow and divide relatively unhindered by mechanical feedback. This accounts for the puzzling result where higher 
overall cell density corresponds to higher cell proliferation. 

\section{Discussion} 
The migratory pattern of motile cells is diverse and depends on  factors such as whether it is a collection of isolated single cells moving in a uniform direction or a collection of adhesive cells which are physically in contact with each other \cite{Lintz2017,Pascalis2017}. 
Here, we present an off-lattice agent-based computational modeling framework for an expanding 2D cell collective. 
By focusing on the influence of noise in the direction of a cell's motion,  
we show that noise strength influences: (i) the migratory pattern and spatial spread or invasion, and (ii) cell density-dependent cell proliferation of cell collectives. 

While the seminal work of Vicsek and co-workers 
has been in many ways foundational to computational modeling-based studies of cell migration \cite{Mehes2014}, few existing models of cell migration consider cell proliferation. Yet, the ability to grow and divide is a fundamental property of many biological systems. 
Our model considers individual cells as active agents that can grow and divide, and whose movement is influenced by their interactions with other cells and stochastic switching in the direction of migration. 
We take into account various biologically relevant inter-cellular interactions, such as cell elastic repulsion, and cell adhesion~\cite{malmi2018cell,sinha2020spatially,sinha2021inter,malmi2022adhesion}. Adhesive interaction between cells, of the type prevalent in confluent tissues, has been taken into account in the past models \cite{Gregoire2003, Peruani2011}.  
The model also includes an additional nearest-neighbor interaction through which cells tend to align the direction of their motion with the average direction of motion of all their neighbors~\cite{vicsek1995novel}. 
Given the recent experimental verification that cell proliferation is pressure-dependent \cite{Bisson2021,di2022pressure}, mechanical feedback on proliferation is an important component of our model as the cell area growth depends on the net force acting on the cell from its contacting neighbors through the $p_c$ term.  
Hence, our model is an important extension of the classical Vicsek model, with self-propelled particles that can undergo growth, birth, and death. 

We find that noise strength strongly influences the migratory pattern of cells in the collective. 
At low noise strengths $\eta=0.01$ and at long times, the cells are sparsely distributed in a ring-like pattern. 
Within this ring, the cells form clusters of different sizes. 
Cells in each of these clusters move in a highly ordered manner, with the orientation of cell velocity aligned in the direction away from the center. 
We quantified this ordered behavior of cell migration in the collective using an order parameter whose value for $\eta=0.01$ is close to 1, indicating a highly ordered motion of cells. 
Cell division events in our model scramble the local order of the cell collective 
as velocity vectors of the daughter cells are assigned random orientations upon division. However, even with these scrambling events present, we notice that the cell collective displays a highly ordered motion at low noise strengths. 
At intermediate noise strengths ($\eta=0.05-0.1$), the spatial distribution of migrating cells still shows a ring-like pattern. Although higher density of cells is still confined to the outer ring, clusters and individual cells are to be found in the interior of this domain as well. 
The orientation order parameter saturates to values much lower than 1 at long times, indicating the onset of a disordered migratory phase. 
The lower value of the order parameter is due to the formation of smaller  cell clusters that move in random directions. 
As the noise strength is further increased to the highest value considered in this study ($\eta=0.2$), we observe a clear change in the migratory pattern and spatial arrangement of cells. 
In this case, higher cell density is observed in the interior of the spatial domain over which cells are distributed. 
The cell collective as a whole is split into multiple smaller clusters, with each cluster moving in random directions. 
The order parameter for the cell collective for such high noise strengths approaches $0$, indicating an almost total loss of orientational order in cell motion. Our results also show that noise strength not only influences the overall spatial pattern but the spread of 
the cell collective as well, which is proportional to the total area covered by the cell collective. The largest spatial spread, compared to the size of the initial distribution of the collective, occurs for very low noise strengths at $\eta=0.01$. In this scenario, cells migrate as a propagating front leading to the emergence of a ring-like pattern. As the noise strength is increased, the spatial spread of the collective is strongly restricted.  

An unexpected result of our study is that noise strength influences cell 
proliferation. Although the total number of cells increases over time for all values 
of noise strength, the trend in proliferation is strongly dependent on the noise strength. 
The total number of cells is almost double the number of cells at the final time point for high noise strength $\eta=0.2$, as compared to $\eta=0.01$. 
Combined with our results showing the effect of noise strength on the spatial spread of the cell collective, we find that cell number density is a highly dynamic 
quantity that increases with noise strength. 
Taken together, we show that as the noise strength increases, the density of the cell collective increases, whereas the orientational order decreases. 

Given the mechanical feedback that limits proliferation due to cell-cell  overlap, the increase of cell number with a higher density is a surprising and counter-intuitive result. 
While the overall density indicates that cells should be more tightly packed at higher noise strengths, our DBSCAN-based cluster analysis shows that the local spatial structure is contrary to what is expected. 
At higher noise strengths, not only do cells form more clusters, but there is a larger number of isolated cells. Isolated cells are ideal sources of proliferation in a collective, characterized by limited mechanical feedback on proliferation from neighboring cells. At lower noise strengths cell clusters contain a larger number of overlapping cells which thus inhibits cell growth and division. In this scenario, cells are localized to the periphery of a ring-like domain while its interior is mostly devoid of cells, leading to the overall density being lower.  
Therefore, even though cell number density is greater at higher noise strengths, there is a larger number of proliferating cells due to the presence of smaller clusters and a greater number of individual cells that are not part of a cluster. 

In conclusion, our study demonstrates that angular fluctuations in cell motility 
direction can strongly determine the spatial distribution of growing cell collectives. 
Our computational model provides a framework for studying the migration of cells in 2D growing cell collectives. 
Our model combines cell velocity re-alignment, as introduced in the Vicsek model, with active growth and cell division.  
This makes our work highly relevant in studying the migration behavior of biological cell collectives, in which cell migration occurs together with cell proliferation. 
Our results imply that there are more, yet unexplained, dynamic behaviors that may emerge from investigating mechanical feedback on proliferation in a system of self-propelled particles undergoing collective motion.\\
\\
\section*{Acknowledgments}
A.M.K acknowledge funding from startup grants. 
The authors acknowledge the support of Augusta University High Performance Computing Services (AUHPCS) for providing computational resources contributing to the results presented in this publication.

\bibliography{library}

\begin{thebibliography}{43}%
\makeatletter
\providecommand \@ifxundefined [1]{%
 \@ifx{#1\undefined}
}%
\providecommand \@ifnum [1]{%
 \ifnum #1\expandafter \@firstoftwo
 \else \expandafter \@secondoftwo
 \fi
}%
\providecommand \@ifx [1]{%
 \ifx #1\expandafter \@firstoftwo
 \else \expandafter \@secondoftwo
 \fi
}%
\providecommand \natexlab [1]{#1}%
\providecommand \enquote  [1]{``#1''}%
\providecommand \bibnamefont  [1]{#1}%
\providecommand \bibfnamefont [1]{#1}%
\providecommand \citenamefont [1]{#1}%
\providecommand \href@noop [0]{\@secondoftwo}%
\providecommand \href [0]{\begingroup \@sanitize@url \@href}%
\providecommand \@href[1]{\@@startlink{#1}\@@href}%
\providecommand \@@href[1]{\endgroup#1\@@endlink}%
\providecommand \@sanitize@url [0]{\catcode `\\12\catcode `\$12\catcode
  `\&12\catcode `\#12\catcode `\^12\catcode `\_12\catcode `\%12\relax}%
\providecommand \@@startlink[1]{}%
\providecommand \@@endlink[0]{}%
\providecommand \url  [0]{\begingroup\@sanitize@url \@url }%
\providecommand \@url [1]{\endgroup\@href {#1}{\urlprefix }}%
\providecommand \urlprefix  [0]{URL }%
\providecommand \Eprint [0]{\href }%
\providecommand \doibase [0]{https://doi.org/}%
\providecommand \selectlanguage [0]{\@gobble}%
\providecommand \bibinfo  [0]{\@secondoftwo}%
\providecommand \bibfield  [0]{\@secondoftwo}%
\providecommand \translation [1]{[#1]}%
\providecommand \BibitemOpen [0]{}%
\providecommand \bibitemStop [0]{}%
\providecommand \bibitemNoStop [0]{.\EOS\space}%
\providecommand \EOS [0]{\spacefactor3000\relax}%
\providecommand \BibitemShut  [1]{\csname bibitem#1\endcsname}%
\let\auto@bib@innerbib\@empty
\bibitem [{\citenamefont {Wu}\ \emph {et~al.}(2012)\citenamefont {Wu},
  \citenamefont {Mao}, \citenamefont {Tan}, \citenamefont {Han}, \citenamefont
  {Ren},\ and\ \citenamefont {Gao}}]{wu2012gradient}%
  \BibitemOpen
  \bibfield  {author} {\bibinfo {author} {\bibfnamefont {J.}~\bibnamefont
  {Wu}}, \bibinfo {author} {\bibfnamefont {Z.}~\bibnamefont {Mao}}, \bibinfo
  {author} {\bibfnamefont {H.}~\bibnamefont {Tan}}, \bibinfo {author}
  {\bibfnamefont {L.}~\bibnamefont {Han}}, \bibinfo {author} {\bibfnamefont
  {T.}~\bibnamefont {Ren}},\ and\ \bibinfo {author} {\bibfnamefont
  {C.}~\bibnamefont {Gao}},\ }\bibfield  {title} {\bibinfo {title} {Gradient
  biomaterials and their influences on cell migration},\ }\href@noop {}
  {\bibfield  {journal} {\bibinfo  {journal} {Interface focus}\ }\textbf
  {\bibinfo {volume} {2}},\ \bibinfo {pages} {337} (\bibinfo {year}
  {2012})}\BibitemShut {NoStop}%
\bibitem [{\citenamefont {Yang}\ \emph {et~al.}(2017)\citenamefont {Yang},
  \citenamefont {Zhao}, \citenamefont {Bai}, \citenamefont {Wang},
  \citenamefont {Tomsia},\ and\ \citenamefont {Bai}}]{yang2017promoting}%
  \BibitemOpen
  \bibfield  {author} {\bibinfo {author} {\bibfnamefont {D.}~\bibnamefont
  {Yang}}, \bibinfo {author} {\bibfnamefont {Z.}~\bibnamefont {Zhao}}, \bibinfo
  {author} {\bibfnamefont {F.}~\bibnamefont {Bai}}, \bibinfo {author}
  {\bibfnamefont {S.}~\bibnamefont {Wang}}, \bibinfo {author} {\bibfnamefont
  {A.~P.}\ \bibnamefont {Tomsia}},\ and\ \bibinfo {author} {\bibfnamefont
  {H.}~\bibnamefont {Bai}},\ }\bibfield  {title} {\bibinfo {title} {Promoting
  cell migration in tissue engineering scaffolds with graded channels},\
  }\href@noop {} {\bibfield  {journal} {\bibinfo  {journal} {Advanced
  Healthcare Materials}\ }\textbf {\bibinfo {volume} {6}},\ \bibinfo {pages}
  {1700472} (\bibinfo {year} {2017})}\BibitemShut {NoStop}%
\bibitem [{\citenamefont {Shim}\ \emph {et~al.}(2021)\citenamefont {Shim},
  \citenamefont {Devenport},\ and\ \citenamefont {Cohen}}]{Shim2021}%
  \BibitemOpen
  \bibfield  {author} {\bibinfo {author} {\bibfnamefont {G.}~\bibnamefont
  {Shim}}, \bibinfo {author} {\bibfnamefont {D.}~\bibnamefont {Devenport}},\
  and\ \bibinfo {author} {\bibfnamefont {D.~J.}\ \bibnamefont {Cohen}},\
  }\bibfield  {title} {\bibinfo {title} {Overriding native cell coordination
  enhances external programming of collective cell migration},\ }\bibfield
  {journal} {\bibinfo  {journal} {Proceedings of the National Academy of
  Sciences of the United States of America}\ }\textbf {\bibinfo {volume}
  {118}},\ \href {https://doi.org/10.1073/pnas.2101352118}
  {10.1073/pnas.2101352118} (\bibinfo {year} {2021})\BibitemShut {NoStop}%
\bibitem [{\citenamefont {Giesen}\ \emph {et~al.}(2014)\citenamefont {Giesen},
  \citenamefont {Wang}, \citenamefont {Schapiro}, \citenamefont {Zivanovic},
  \citenamefont {Jacobs}, \citenamefont {Hattendorf}, \citenamefont
  {Sch{\"u}ffler}, \citenamefont {Grolimund}, \citenamefont {Buhmann},
  \citenamefont {Brandt} \emph {et~al.}}]{giesen2014highly}%
  \BibitemOpen
  \bibfield  {author} {\bibinfo {author} {\bibfnamefont {C.}~\bibnamefont
  {Giesen}}, \bibinfo {author} {\bibfnamefont {H.~A.}\ \bibnamefont {Wang}},
  \bibinfo {author} {\bibfnamefont {D.}~\bibnamefont {Schapiro}}, \bibinfo
  {author} {\bibfnamefont {N.}~\bibnamefont {Zivanovic}}, \bibinfo {author}
  {\bibfnamefont {A.}~\bibnamefont {Jacobs}}, \bibinfo {author} {\bibfnamefont
  {B.}~\bibnamefont {Hattendorf}}, \bibinfo {author} {\bibfnamefont {P.~J.}\
  \bibnamefont {Sch{\"u}ffler}}, \bibinfo {author} {\bibfnamefont
  {D.}~\bibnamefont {Grolimund}}, \bibinfo {author} {\bibfnamefont {J.~M.}\
  \bibnamefont {Buhmann}}, \bibinfo {author} {\bibfnamefont {S.}~\bibnamefont
  {Brandt}}, \emph {et~al.},\ }\bibfield  {title} {\bibinfo {title} {Highly
  multiplexed imaging of tumor tissues with subcellular resolution by mass
  cytometry},\ }\href@noop {} {\bibfield  {journal} {\bibinfo  {journal}
  {Nature methods}\ }\textbf {\bibinfo {volume} {11}},\ \bibinfo {pages} {417}
  (\bibinfo {year} {2014})}\BibitemShut {NoStop}%
\bibitem [{\citenamefont {Lin}\ \emph {et~al.}(2018)\citenamefont {Lin},
  \citenamefont {Izar}, \citenamefont {Wang}, \citenamefont {Yapp},
  \citenamefont {Mei}, \citenamefont {Shah}, \citenamefont {Santagata},\ and\
  \citenamefont {Sorger}}]{lin2018highly}%
  \BibitemOpen
  \bibfield  {author} {\bibinfo {author} {\bibfnamefont {J.-R.}\ \bibnamefont
  {Lin}}, \bibinfo {author} {\bibfnamefont {B.}~\bibnamefont {Izar}}, \bibinfo
  {author} {\bibfnamefont {S.}~\bibnamefont {Wang}}, \bibinfo {author}
  {\bibfnamefont {C.}~\bibnamefont {Yapp}}, \bibinfo {author} {\bibfnamefont
  {S.}~\bibnamefont {Mei}}, \bibinfo {author} {\bibfnamefont {P.~M.}\
  \bibnamefont {Shah}}, \bibinfo {author} {\bibfnamefont {S.}~\bibnamefont
  {Santagata}},\ and\ \bibinfo {author} {\bibfnamefont {P.~K.}\ \bibnamefont
  {Sorger}},\ }\bibfield  {title} {\bibinfo {title} {Highly multiplexed
  immunofluorescence imaging of human tissues and tumors using t-cycif and
  conventional optical microscopes},\ }\href@noop {} {\bibfield  {journal}
  {\bibinfo  {journal} {Elife}\ }\textbf {\bibinfo {volume} {7}} (\bibinfo
  {year} {2018})}\BibitemShut {NoStop}%
\bibitem [{\citenamefont {Gaglia}\ \emph {et~al.}(2022)\citenamefont {Gaglia},
  \citenamefont {Kabraji}, \citenamefont {Rammos}, \citenamefont {Dai},
  \citenamefont {Verma}, \citenamefont {Wang}, \citenamefont {Mills},
  \citenamefont {Chung}, \citenamefont {Bergholz}, \citenamefont {Coy} \emph
  {et~al.}}]{gaglia2022temporal}%
  \BibitemOpen
  \bibfield  {author} {\bibinfo {author} {\bibfnamefont {G.}~\bibnamefont
  {Gaglia}}, \bibinfo {author} {\bibfnamefont {S.}~\bibnamefont {Kabraji}},
  \bibinfo {author} {\bibfnamefont {D.}~\bibnamefont {Rammos}}, \bibinfo
  {author} {\bibfnamefont {Y.}~\bibnamefont {Dai}}, \bibinfo {author}
  {\bibfnamefont {A.}~\bibnamefont {Verma}}, \bibinfo {author} {\bibfnamefont
  {S.}~\bibnamefont {Wang}}, \bibinfo {author} {\bibfnamefont {C.~E.}\
  \bibnamefont {Mills}}, \bibinfo {author} {\bibfnamefont {M.}~\bibnamefont
  {Chung}}, \bibinfo {author} {\bibfnamefont {J.~S.}\ \bibnamefont {Bergholz}},
  \bibinfo {author} {\bibfnamefont {S.}~\bibnamefont {Coy}}, \emph {et~al.},\
  }\bibfield  {title} {\bibinfo {title} {Temporal and spatial topography of
  cell proliferation in cancer},\ }\href@noop {} {\bibfield  {journal}
  {\bibinfo  {journal} {Nature Cell Biology}\ }\textbf {\bibinfo {volume}
  {24}},\ \bibinfo {pages} {316} (\bibinfo {year} {2022})}\BibitemShut
  {NoStop}%
\bibitem [{\citenamefont {Shraiman}(2005)}]{shraiman2005mechanical}%
  \BibitemOpen
  \bibfield  {author} {\bibinfo {author} {\bibfnamefont {B.~I.}\ \bibnamefont
  {Shraiman}},\ }\bibfield  {title} {\bibinfo {title} {Mechanical feedback as a
  possible regulator of tissue growth},\ }\href@noop {} {\bibfield  {journal}
  {\bibinfo  {journal} {Proceedings of the National Academy of Sciences}\
  }\textbf {\bibinfo {volume} {102}},\ \bibinfo {pages} {3318} (\bibinfo {year}
  {2005})}\BibitemShut {NoStop}%
\bibitem [{\citenamefont {Streichan}\ \emph {et~al.}(2014)\citenamefont
  {Streichan}, \citenamefont {Hoerner}, \citenamefont {Schneidt}, \citenamefont
  {Holzer},\ and\ \citenamefont {Hufnagel}}]{streichan2014spatial}%
  \BibitemOpen
  \bibfield  {author} {\bibinfo {author} {\bibfnamefont {S.~J.}\ \bibnamefont
  {Streichan}}, \bibinfo {author} {\bibfnamefont {C.~R.}\ \bibnamefont
  {Hoerner}}, \bibinfo {author} {\bibfnamefont {T.}~\bibnamefont {Schneidt}},
  \bibinfo {author} {\bibfnamefont {D.}~\bibnamefont {Holzer}},\ and\ \bibinfo
  {author} {\bibfnamefont {L.}~\bibnamefont {Hufnagel}},\ }\bibfield  {title}
  {\bibinfo {title} {Spatial constraints control cell proliferation in
  tissues},\ }\href@noop {} {\bibfield  {journal} {\bibinfo  {journal}
  {Proceedings of the National Academy of Sciences}\ }\textbf {\bibinfo
  {volume} {111}},\ \bibinfo {pages} {5586} (\bibinfo {year}
  {2014})}\BibitemShut {NoStop}%
\bibitem [{\citenamefont {Malmi-Kakkada}\ \emph {et~al.}(2022)\citenamefont
  {Malmi-Kakkada}, \citenamefont {Sinha}, \citenamefont {Li},\ and\
  \citenamefont {Thirumalai}}]{malmi2022adhesion}%
  \BibitemOpen
  \bibfield  {author} {\bibinfo {author} {\bibfnamefont {A.~N.}\ \bibnamefont
  {Malmi-Kakkada}}, \bibinfo {author} {\bibfnamefont {S.}~\bibnamefont
  {Sinha}}, \bibinfo {author} {\bibfnamefont {X.}~\bibnamefont {Li}},\ and\
  \bibinfo {author} {\bibfnamefont {D.}~\bibnamefont {Thirumalai}},\ }\bibfield
   {title} {\bibinfo {title} {Adhesion strength between cells regulate
  non-monotonic growth by a biomechanical feedback mechanism},\ }\href@noop {}
  {\bibfield  {journal} {\bibinfo  {journal} {Biophysical Journal}\ }\textbf
  {\bibinfo {volume} {121}},\ \bibinfo {pages} {3719} (\bibinfo {year}
  {2022})}\BibitemShut {NoStop}%
\bibitem [{\citenamefont {Ladoux}\ and\ \citenamefont
  {Mège}(2017)}]{Ladoux2017}%
  \BibitemOpen
  \bibfield  {author} {\bibinfo {author} {\bibfnamefont {B.}~\bibnamefont
  {Ladoux}}\ and\ \bibinfo {author} {\bibfnamefont {R.~M.}\ \bibnamefont
  {Mège}},\ }\bibfield  {title} {\bibinfo {title} {Mechanobiology of
  collective cell behaviours},\ }\bibfield  {journal} {\bibinfo  {journal}
  {Nature Reviews Molecular Cell Biology}\ }\textbf {\bibinfo {volume} {18}},\
  \href {https://doi.org/10.1038/nrm.2017.98} {10.1038/nrm.2017.98} (\bibinfo
  {year} {2017})\BibitemShut {NoStop}%
\bibitem [{\citenamefont {Angelini}\ \emph {et~al.}(2010)\citenamefont
  {Angelini}, \citenamefont {Hannezo}, \citenamefont {Trepat}, \citenamefont
  {Fredberg},\ and\ \citenamefont {Weitz}}]{Angelini2010}%
  \BibitemOpen
  \bibfield  {author} {\bibinfo {author} {\bibfnamefont {T.~E.}\ \bibnamefont
  {Angelini}}, \bibinfo {author} {\bibfnamefont {E.}~\bibnamefont {Hannezo}},
  \bibinfo {author} {\bibfnamefont {X.}~\bibnamefont {Trepat}}, \bibinfo
  {author} {\bibfnamefont {J.~J.}\ \bibnamefont {Fredberg}},\ and\ \bibinfo
  {author} {\bibfnamefont {D.~A.}\ \bibnamefont {Weitz}},\ }\bibfield  {title}
  {\bibinfo {title} {Cell migration driven by cooperative substrate deformation
  patterns},\ }\bibfield  {journal} {\bibinfo  {journal} {Physical Review
  Letters}\ }\textbf {\bibinfo {volume} {104}},\ \href
  {https://doi.org/10.1103/PhysRevLett.104.168104}
  {10.1103/PhysRevLett.104.168104} (\bibinfo {year} {2010})\BibitemShut
  {NoStop}%
\bibitem [{\citenamefont {Pascalis}\ and\ \citenamefont
  {Etienne-Manneville}(2017)}]{Pascalis2017}%
  \BibitemOpen
  \bibfield  {author} {\bibinfo {author} {\bibfnamefont {C.~D.}\ \bibnamefont
  {Pascalis}}\ and\ \bibinfo {author} {\bibfnamefont {S.}~\bibnamefont
  {Etienne-Manneville}},\ }\bibfield  {title} {\bibinfo {title} {Single and
  collective cell migration: The mechanics of adhesions},\ }\bibfield
  {journal} {\bibinfo  {journal} {Molecular Biology of the Cell}\ }\textbf
  {\bibinfo {volume} {28}},\ \href {https://doi.org/10.1091/mbc.E17-03-0134}
  {10.1091/mbc.E17-03-0134} (\bibinfo {year} {2017})\BibitemShut {NoStop}%
\bibitem [{\citenamefont {Méhes}\ and\ \citenamefont
  {Vicsek}(2014)}]{Mehes2014}%
  \BibitemOpen
  \bibfield  {author} {\bibinfo {author} {\bibfnamefont {E.}~\bibnamefont
  {Méhes}}\ and\ \bibinfo {author} {\bibfnamefont {T.}~\bibnamefont
  {Vicsek}},\ }\bibfield  {title} {\bibinfo {title} {Collective motion of
  cells: From experiments to models},\ }\bibfield  {journal} {\bibinfo
  {journal} {Integrative Biology (United Kingdom)}\ }\textbf {\bibinfo {volume}
  {6}},\ \href {https://doi.org/10.1039/c4ib00115j} {10.1039/c4ib00115j}
  (\bibinfo {year} {2014})\BibitemShut {NoStop}%
\bibitem [{\citenamefont {Kuriyama}\ \emph {et~al.}(2014)\citenamefont
  {Kuriyama}, \citenamefont {Theveneau}, \citenamefont {Benedetto},
  \citenamefont {Parsons}, \citenamefont {Tanaka}, \citenamefont {Charras},
  \citenamefont {Kabla},\ and\ \citenamefont {Mayor}}]{Kuriyama2014}%
  \BibitemOpen
  \bibfield  {author} {\bibinfo {author} {\bibfnamefont {S.}~\bibnamefont
  {Kuriyama}}, \bibinfo {author} {\bibfnamefont {E.}~\bibnamefont {Theveneau}},
  \bibinfo {author} {\bibfnamefont {A.}~\bibnamefont {Benedetto}}, \bibinfo
  {author} {\bibfnamefont {M.}~\bibnamefont {Parsons}}, \bibinfo {author}
  {\bibfnamefont {M.}~\bibnamefont {Tanaka}}, \bibinfo {author} {\bibfnamefont
  {G.}~\bibnamefont {Charras}}, \bibinfo {author} {\bibfnamefont
  {A.}~\bibnamefont {Kabla}},\ and\ \bibinfo {author} {\bibfnamefont
  {R.}~\bibnamefont {Mayor}},\ }\bibfield  {title} {\bibinfo {title} {In vivo
  collective cell migration requires an lpar2-dependent increase in tissue
  fluidity},\ }\bibfield  {journal} {\bibinfo  {journal} {Journal of Cell
  Biology}\ }\textbf {\bibinfo {volume} {206}},\ \href
  {https://doi.org/10.1083/jcb.201402093} {10.1083/jcb.201402093} (\bibinfo
  {year} {2014})\BibitemShut {NoStop}%
\bibitem [{\citenamefont {Huang}\ \emph {et~al.}(2004)\citenamefont {Huang},
  \citenamefont {Jacobson},\ and\ \citenamefont {Schaller}}]{Huang2004}%
  \BibitemOpen
  \bibfield  {author} {\bibinfo {author} {\bibfnamefont {C.}~\bibnamefont
  {Huang}}, \bibinfo {author} {\bibfnamefont {K.}~\bibnamefont {Jacobson}},\
  and\ \bibinfo {author} {\bibfnamefont {M.~D.}\ \bibnamefont {Schaller}},\
  }\bibfield  {title} {\bibinfo {title} {Map kinases and cell migration},\
  }\bibfield  {journal} {\bibinfo  {journal} {Journal of Cell Science}\
  }\textbf {\bibinfo {volume} {117}},\ \href
  {https://doi.org/10.1242/jcs.01481} {10.1242/jcs.01481} (\bibinfo {year}
  {2004})\BibitemShut {NoStop}%
\bibitem [{\citenamefont {Melani}\ \emph {et~al.}(2008)\citenamefont {Melani},
  \citenamefont {Simpson}, \citenamefont {Brugge},\ and\ \citenamefont
  {Montell}}]{Melani2008}%
  \BibitemOpen
  \bibfield  {author} {\bibinfo {author} {\bibfnamefont {M.}~\bibnamefont
  {Melani}}, \bibinfo {author} {\bibfnamefont {K.~J.}\ \bibnamefont {Simpson}},
  \bibinfo {author} {\bibfnamefont {J.~S.}\ \bibnamefont {Brugge}},\ and\
  \bibinfo {author} {\bibfnamefont {D.}~\bibnamefont {Montell}},\ }\bibfield
  {title} {\bibinfo {title} {Regulation of cell adhesion and collective cell
  migration by hindsight and its human homolog rreb1},\ }\bibfield  {journal}
  {\bibinfo  {journal} {Current Biology}\ }\textbf {\bibinfo {volume} {18}},\
  \href {https://doi.org/10.1016/j.cub.2008.03.024} {10.1016/j.cub.2008.03.024}
  (\bibinfo {year} {2008})\BibitemShut {NoStop}%
\bibitem [{\citenamefont {Malmi-Kakkada}\ \emph
  {et~al.}(2018{\natexlab{a}})\citenamefont {Malmi-Kakkada}, \citenamefont
  {Li}, \citenamefont {Samanta}, \citenamefont {Sinha},\ and\ \citenamefont
  {Thirumalai}}]{malmi2018cell}%
  \BibitemOpen
  \bibfield  {author} {\bibinfo {author} {\bibfnamefont {A.~N.}\ \bibnamefont
  {Malmi-Kakkada}}, \bibinfo {author} {\bibfnamefont {X.}~\bibnamefont {Li}},
  \bibinfo {author} {\bibfnamefont {H.~S.}\ \bibnamefont {Samanta}}, \bibinfo
  {author} {\bibfnamefont {S.}~\bibnamefont {Sinha}},\ and\ \bibinfo {author}
  {\bibfnamefont {D.}~\bibnamefont {Thirumalai}},\ }\bibfield  {title}
  {\bibinfo {title} {Cell growth rate dictates the onset of glass to fluidlike
  transition and long time superdiffusion in an evolving cell colony},\
  }\href@noop {} {\bibfield  {journal} {\bibinfo  {journal} {Physical Review
  X}\ }\textbf {\bibinfo {volume} {8}},\ \bibinfo {pages} {021025} (\bibinfo
  {year} {2018}{\natexlab{a}})}\BibitemShut {NoStop}%
\bibitem [{\citenamefont {Sinha}\ \emph {et~al.}(2020)\citenamefont {Sinha},
  \citenamefont {Malmi-Kakkada}, \citenamefont {Li}, \citenamefont {Samanta},\
  and\ \citenamefont {Thirumalai}}]{sinha2020spatially}%
  \BibitemOpen
  \bibfield  {author} {\bibinfo {author} {\bibfnamefont {S.}~\bibnamefont
  {Sinha}}, \bibinfo {author} {\bibfnamefont {A.~N.}\ \bibnamefont
  {Malmi-Kakkada}}, \bibinfo {author} {\bibfnamefont {X.}~\bibnamefont {Li}},
  \bibinfo {author} {\bibfnamefont {H.~S.}\ \bibnamefont {Samanta}},\ and\
  \bibinfo {author} {\bibfnamefont {D.}~\bibnamefont {Thirumalai}},\ }\bibfield
   {title} {\bibinfo {title} {Spatially heterogeneous dynamics of cells in a
  growing tumor spheroid: Comparison between theory and experiments},\
  }\href@noop {} {\bibfield  {journal} {\bibinfo  {journal} {Soft Matter}\
  }\textbf {\bibinfo {volume} {16}},\ \bibinfo {pages} {5294} (\bibinfo {year}
  {2020})}\BibitemShut {NoStop}%
\bibitem [{\citenamefont {Sinha}\ and\ \citenamefont
  {Malmi-Kakkada}(2021)}]{sinha2021inter}%
  \BibitemOpen
  \bibfield  {author} {\bibinfo {author} {\bibfnamefont {S.}~\bibnamefont
  {Sinha}}\ and\ \bibinfo {author} {\bibfnamefont {A.~N.}\ \bibnamefont
  {Malmi-Kakkada}},\ }\bibfield  {title} {\bibinfo {title} {Inter-particle
  adhesion regulates the surface roughness of growing dense three-dimensional
  active particle aggregates},\ }\href@noop {} {\bibfield  {journal} {\bibinfo
  {journal} {J. Phys. Chem. B}\ }\textbf {\bibinfo {volume} {125}},\ \bibinfo
  {pages} {10445} (\bibinfo {year} {2021})}\BibitemShut {NoStop}%
\bibitem [{\citenamefont {JE}\ \emph {et~al.}(2022)\citenamefont {JE},
  \citenamefont {T}, \citenamefont {K}, \citenamefont {R}, \citenamefont {van
  Rienen~U}, \citenamefont {R},\ and\ \citenamefont {R}}]{Dawson2022b}%
  \BibitemOpen
  \bibfield  {author} {\bibinfo {author} {\bibfnamefont {D.}~\bibnamefont
  {JE}}, \bibinfo {author} {\bibfnamefont {S.}~\bibnamefont {T}}, \bibinfo
  {author} {\bibfnamefont {P.}~\bibnamefont {K}}, \bibinfo {author}
  {\bibfnamefont {B.}~\bibnamefont {R}}, \bibinfo {author} {\bibnamefont {van
  Rienen~U}}, \bibinfo {author} {\bibfnamefont {A.}~\bibnamefont {R}},\ and\
  \bibinfo {author} {\bibfnamefont {K.}~\bibnamefont {R}},\ }\bibfield  {title}
  {\bibinfo {title} {Cell-cell interactions and fluctuations in the direction
  of motility promote directed migration of osteoblasts in direct current
  electrotaxis.},\ }\bibfield  {journal} {\bibinfo  {journal} {Frontiers in
  Bioengineering and Biotechnology}\ }\href
  {https://doi.org/10.3389/fbioe.2022.995326} {10.3389/fbioe.2022.995326}
  (\bibinfo {year} {2022})\BibitemShut {NoStop}%
\bibitem [{\citenamefont {Yang}\ \emph {et~al.}(2014)\citenamefont {Yang},
  \citenamefont {Manning},\ and\ \citenamefont {Marchetti}}]{Yang2014}%
  \BibitemOpen
  \bibfield  {author} {\bibinfo {author} {\bibfnamefont {X.}~\bibnamefont
  {Yang}}, \bibinfo {author} {\bibfnamefont {M.~L.}\ \bibnamefont {Manning}},\
  and\ \bibinfo {author} {\bibfnamefont {M.~C.}\ \bibnamefont {Marchetti}},\
  }\bibfield  {title} {\bibinfo {title} {Aggregation and segregation of
  confined active particles},\ }\bibfield  {journal} {\bibinfo  {journal} {Soft
  Matter}\ }\textbf {\bibinfo {volume} {10}},\ \href
  {https://doi.org/10.1039/c4sm00927d} {10.1039/c4sm00927d} (\bibinfo {year}
  {2014})\BibitemShut {NoStop}%
\bibitem [{\citenamefont {Woods}\ \emph {et~al.}(2014)\citenamefont {Woods},
  \citenamefont {Carmona-Fontaine}, \citenamefont {Barnes}, \citenamefont
  {Couzin}, \citenamefont {Mayor},\ and\ \citenamefont {Page}}]{Woods2014}%
  \BibitemOpen
  \bibfield  {author} {\bibinfo {author} {\bibfnamefont {M.~L.}\ \bibnamefont
  {Woods}}, \bibinfo {author} {\bibfnamefont {C.}~\bibnamefont
  {Carmona-Fontaine}}, \bibinfo {author} {\bibfnamefont {C.~P.}\ \bibnamefont
  {Barnes}}, \bibinfo {author} {\bibfnamefont {I.~D.}\ \bibnamefont {Couzin}},
  \bibinfo {author} {\bibfnamefont {R.}~\bibnamefont {Mayor}},\ and\ \bibinfo
  {author} {\bibfnamefont {K.~M.}\ \bibnamefont {Page}},\ }\bibfield  {title}
  {\bibinfo {title} {Directional collective cell migration emerges as a
  property of cell interactions},\ }\bibfield  {journal} {\bibinfo  {journal}
  {PLoS ONE}\ }\textbf {\bibinfo {volume} {9}},\ \href
  {https://doi.org/10.1371/journal.pone.0104969} {10.1371/journal.pone.0104969}
  (\bibinfo {year} {2014})\BibitemShut {NoStop}%
\bibitem [{\citenamefont {Battersby}(2015)}]{Battersby2015}%
  \BibitemOpen
  \bibfield  {author} {\bibinfo {author} {\bibfnamefont {S.}~\bibnamefont
  {Battersby}},\ }\bibfield  {title} {\bibinfo {title} {News feature: The cells
  that flock together},\ }\href@noop {} {\bibfield  {journal} {\bibinfo
  {journal} {Proceedings of the National Academy of Sciences}\ }\textbf
  {\bibinfo {volume} {112}} (\bibinfo {year} {2015})}\BibitemShut {NoStop}%
\bibitem [{\citenamefont {Grégoire}\ \emph {et~al.}(2003)\citenamefont
  {Grégoire}, \citenamefont {Chaté},\ and\ \citenamefont
  {Tu}}]{Gregoire2003}%
  \BibitemOpen
  \bibfield  {author} {\bibinfo {author} {\bibfnamefont {G.}~\bibnamefont
  {Grégoire}}, \bibinfo {author} {\bibfnamefont {H.}~\bibnamefont {Chaté}},\
  and\ \bibinfo {author} {\bibfnamefont {Y.}~\bibnamefont {Tu}},\ }\bibfield
  {title} {\bibinfo {title} {Moving and staying together without a leader},\
  }\bibfield  {journal} {\bibinfo  {journal} {Physica D: Nonlinear Phenomena}\
  }\textbf {\bibinfo {volume} {181}},\ \href
  {https://doi.org/10.1016/S0167-2789(03)00102-7}
  {10.1016/S0167-2789(03)00102-7} (\bibinfo {year} {2003})\BibitemShut
  {NoStop}%
\bibitem [{\citenamefont {Buttensch{\"o}n}\ and\ \citenamefont
  {Edelstein-Keshet}(2020)}]{Buttenschon2020}%
  \BibitemOpen
  \bibfield  {author} {\bibinfo {author} {\bibfnamefont {A.}~\bibnamefont
  {Buttensch{\"o}n}}\ and\ \bibinfo {author} {\bibfnamefont {L.}~\bibnamefont
  {Edelstein-Keshet}},\ }\bibfield  {title} {\bibinfo {title} {Bridging from
  single to collective cell migration: A review of models and links to
  experiments},\ }\href {https://doi.org/10.1371/journal.pcbi.1008411}
  {\bibfield  {journal} {\bibinfo  {journal} {PLOS Computational Biology}\
  }\textbf {\bibinfo {volume} {16}},\ \bibinfo {pages} {1} (\bibinfo {year}
  {2020})}\BibitemShut {NoStop}%
\bibitem [{\citenamefont {Huebner}\ \emph {et~al.}(2021)\citenamefont
  {Huebner}, \citenamefont {Malmi-Kakkada}, \citenamefont {Sarikaya},
  \citenamefont {Weng}, \citenamefont {Thirumalai},\ and\ \citenamefont
  {Wallingford}}]{huebner2021mechanical}%
  \BibitemOpen
  \bibfield  {author} {\bibinfo {author} {\bibfnamefont {R.~J.}\ \bibnamefont
  {Huebner}}, \bibinfo {author} {\bibfnamefont {A.~N.}\ \bibnamefont
  {Malmi-Kakkada}}, \bibinfo {author} {\bibfnamefont {S.}~\bibnamefont
  {Sarikaya}}, \bibinfo {author} {\bibfnamefont {S.}~\bibnamefont {Weng}},
  \bibinfo {author} {\bibfnamefont {D.}~\bibnamefont {Thirumalai}},\ and\
  \bibinfo {author} {\bibfnamefont {J.~B.}\ \bibnamefont {Wallingford}},\
  }\bibfield  {title} {\bibinfo {title} {Mechanical heterogeneity along single
  cell-cell junctions is driven by lateral clustering of cadherins during
  vertebrate axis elongation},\ }\href@noop {} {\bibfield  {journal} {\bibinfo
  {journal} {Elife}\ }\textbf {\bibinfo {volume} {10}},\ \bibinfo {pages}
  {e65390} (\bibinfo {year} {2021})}\BibitemShut {NoStop}%
\bibitem [{\citenamefont {Vicsek}\ \emph {et~al.}(1995)\citenamefont {Vicsek},
  \citenamefont {Czir{\'o}k}, \citenamefont {Ben-Jacob}, \citenamefont
  {Cohen},\ and\ \citenamefont {Shochet}}]{vicsek1995novel}%
  \BibitemOpen
  \bibfield  {author} {\bibinfo {author} {\bibfnamefont {T.}~\bibnamefont
  {Vicsek}}, \bibinfo {author} {\bibfnamefont {A.}~\bibnamefont {Czir{\'o}k}},
  \bibinfo {author} {\bibfnamefont {E.}~\bibnamefont {Ben-Jacob}}, \bibinfo
  {author} {\bibfnamefont {I.}~\bibnamefont {Cohen}},\ and\ \bibinfo {author}
  {\bibfnamefont {O.}~\bibnamefont {Shochet}},\ }\bibfield  {title} {\bibinfo
  {title} {Novel type of phase transition in a system of self-driven
  particles},\ }\href@noop {} {\bibfield  {journal} {\bibinfo  {journal}
  {Physical review letters}\ }\textbf {\bibinfo {volume} {75}},\ \bibinfo
  {pages} {1226} (\bibinfo {year} {1995})}\BibitemShut {NoStop}%
\bibitem [{\citenamefont {Binny}\ \emph {et~al.}(2015)\citenamefont {Binny},
  \citenamefont {Plank},\ and\ \citenamefont {James}}]{Binny2015}%
  \BibitemOpen
  \bibfield  {author} {\bibinfo {author} {\bibfnamefont {R.~N.}\ \bibnamefont
  {Binny}}, \bibinfo {author} {\bibfnamefont {M.~J.}\ \bibnamefont {Plank}},\
  and\ \bibinfo {author} {\bibfnamefont {A.}~\bibnamefont {James}},\ }\bibfield
   {title} {\bibinfo {title} {Spatial moment dynamics for collective cell
  movement incorporating a neighbour-dependent directional bias},\ }\bibfield
  {journal} {\bibinfo  {journal} {Journal of the Royal Society Interface}\
  }\textbf {\bibinfo {volume} {12}},\ \href
  {https://doi.org/10.1098/rsif.2015.0228} {10.1098/rsif.2015.0228} (\bibinfo
  {year} {2015})\BibitemShut {NoStop}%
\bibitem [{\citenamefont {Pawan K.~Mishra}(2022)}]{Mishra2022}%
  \BibitemOpen
  \bibfield  {author} {\bibinfo {author} {\bibfnamefont {S.~M.}\ \bibnamefont
  {Pawan K.~Mishra}},\ }\bibfield  {title} {\bibinfo {title} {Active polar
  flock with birth and death},\ }\bibfield  {journal} {\bibinfo  {journal}
  {Physics of Fluids}\ }\href
  {https://doi.org/https://doi.org/10.1063/5.0086952}
  {https://doi.org/10.1063/5.0086952} (\bibinfo {year} {2022})\BibitemShut
  {NoStop}%
\bibitem [{\citenamefont {Heinrich}\ \emph {et~al.}(2020)\citenamefont
  {Heinrich}, \citenamefont {Alert}, \citenamefont {LaChance}, \citenamefont
  {Zajdel}, \citenamefont {Ko{\v s}mrlj},\ and\ \citenamefont
  {Cohen}}]{Heinrich2020}%
  \BibitemOpen
  \bibfield  {author} {\bibinfo {author} {\bibfnamefont {M.~A.}\ \bibnamefont
  {Heinrich}}, \bibinfo {author} {\bibfnamefont {R.}~\bibnamefont {Alert}},
  \bibinfo {author} {\bibfnamefont {J.~M.}\ \bibnamefont {LaChance}}, \bibinfo
  {author} {\bibfnamefont {T.~J.}\ \bibnamefont {Zajdel}}, \bibinfo {author}
  {\bibfnamefont {A.}~\bibnamefont {Ko{\v s}mrlj}},\ and\ \bibinfo {author}
  {\bibfnamefont {D.~J.}\ \bibnamefont {Cohen}},\ }\bibfield  {title} {\bibinfo
  {title} {Size-dependent patterns of cell proliferation and migration in
  freely-expanding epithelia},\ }\href {https://doi.org/10.7554/eLife.58945}
  {\bibfield  {journal} {\bibinfo  {journal} {eLife}\ }\textbf {\bibinfo
  {volume} {9}},\ \bibinfo {pages} {e58945} (\bibinfo {year}
  {2020})}\BibitemShut {NoStop}%
\bibitem [{\citenamefont {Malmi-Kakkada}\ \emph
  {et~al.}(2018{\natexlab{b}})\citenamefont {Malmi-Kakkada}, \citenamefont
  {Li}, \citenamefont {Samanta}, \citenamefont {Sinha},\ and\ \citenamefont
  {Thirumalai}}]{Malmi-Kakkada2018}%
  \BibitemOpen
  \bibfield  {author} {\bibinfo {author} {\bibfnamefont {A.~N.}\ \bibnamefont
  {Malmi-Kakkada}}, \bibinfo {author} {\bibfnamefont {X.}~\bibnamefont {Li}},
  \bibinfo {author} {\bibfnamefont {H.~S.}\ \bibnamefont {Samanta}}, \bibinfo
  {author} {\bibfnamefont {S.}~\bibnamefont {Sinha}},\ and\ \bibinfo {author}
  {\bibfnamefont {D.}~\bibnamefont {Thirumalai}},\ }\bibfield  {title}
  {\bibinfo {title} {Cell growth rate dictates the onset of glass to fluidlike
  transition and long time superdiffusion in an evolving cell colony},\ }\href
  {https://doi.org/10.1103/PhysRevX.8.021025} {\bibfield  {journal} {\bibinfo
  {journal} {Phys. Rev. X}\ }\textbf {\bibinfo {volume} {8}},\ \bibinfo {pages}
  {021025} (\bibinfo {year} {2018}{\natexlab{b}})}\BibitemShut {NoStop}%
\bibitem [{\citenamefont {Di~Meglio}\ \emph {et~al.}(2022)\citenamefont
  {Di~Meglio}, \citenamefont {Trushko}, \citenamefont {Guillamat},
  \citenamefont {Blanch-Mercader}, \citenamefont {Abuhattum},\ and\
  \citenamefont {Roux}}]{di2022pressure}%
  \BibitemOpen
  \bibfield  {author} {\bibinfo {author} {\bibfnamefont {I.}~\bibnamefont
  {Di~Meglio}}, \bibinfo {author} {\bibfnamefont {A.}~\bibnamefont {Trushko}},
  \bibinfo {author} {\bibfnamefont {P.}~\bibnamefont {Guillamat}}, \bibinfo
  {author} {\bibfnamefont {C.}~\bibnamefont {Blanch-Mercader}}, \bibinfo
  {author} {\bibfnamefont {S.}~\bibnamefont {Abuhattum}},\ and\ \bibinfo
  {author} {\bibfnamefont {A.}~\bibnamefont {Roux}},\ }\bibfield  {title}
  {\bibinfo {title} {Pressure and curvature control of the cell cycle in
  epithelia growing under spherical confinement},\ }\href@noop {} {\bibfield
  {journal} {\bibinfo  {journal} {Cell reports}\ }\textbf {\bibinfo {volume}
  {40}},\ \bibinfo {pages} {111227} (\bibinfo {year} {2022})}\BibitemShut
  {NoStop}%
\bibitem [{\citenamefont {Zills}\ \emph {et~al.}(2023)\citenamefont {Zills},
  \citenamefont {Datta},\ and\ \citenamefont
  {Malmi-Kakkada}}]{zills2023enhanced}%
  \BibitemOpen
  \bibfield  {author} {\bibinfo {author} {\bibfnamefont {G.}~\bibnamefont
  {Zills}}, \bibinfo {author} {\bibfnamefont {T.}~\bibnamefont {Datta}},\ and\
  \bibinfo {author} {\bibfnamefont {A.~N.}\ \bibnamefont {Malmi-Kakkada}},\
  }\bibfield  {title} {\bibinfo {title} {Enhanced mechanical heterogeneity of
  cell collectives due to temporal fluctuations in cell elasticity},\
  }\href@noop {} {\bibfield  {journal} {\bibinfo  {journal} {Physical Review
  E}\ }\textbf {\bibinfo {volume} {107}},\ \bibinfo {pages} {014401} (\bibinfo
  {year} {2023})}\BibitemShut {NoStop}%
\bibitem [{\citenamefont {Drasdo}\ and\ \citenamefont
  {H{\"o}hme}(2005)}]{drasdo2005single}%
  \BibitemOpen
  \bibfield  {author} {\bibinfo {author} {\bibfnamefont {D.}~\bibnamefont
  {Drasdo}}\ and\ \bibinfo {author} {\bibfnamefont {S.}~\bibnamefont
  {H{\"o}hme}},\ }\bibfield  {title} {\bibinfo {title} {A single-cell-based
  model of tumor growth in vitro: monolayers and spheroids},\ }\href@noop {}
  {\bibfield  {journal} {\bibinfo  {journal} {Phys Biol}\ }\textbf {\bibinfo
  {volume} {2}},\ \bibinfo {pages} {3} (\bibinfo {year} {2005})}\BibitemShut
  {NoStop}%
\bibitem [{\citenamefont {Schaller}\ and\ \citenamefont
  {Meyer-Hermann}(2005)}]{schaller2005multicellular}%
  \BibitemOpen
  \bibfield  {author} {\bibinfo {author} {\bibfnamefont {G.}~\bibnamefont
  {Schaller}}\ and\ \bibinfo {author} {\bibfnamefont {M.}~\bibnamefont
  {Meyer-Hermann}},\ }\bibfield  {title} {\bibinfo {title} {Multicellular tumor
  spheroid in an off-lattice voronoi-delaunay cell model},\ }\href@noop {}
  {\bibfield  {journal} {\bibinfo  {journal} {Physical Review E}\ }\textbf
  {\bibinfo {volume} {71}},\ \bibinfo {pages} {051910} (\bibinfo {year}
  {2005})}\BibitemShut {NoStop}%
\bibitem [{\citenamefont {Shamir}\ \emph {et~al.}(2016)\citenamefont {Shamir},
  \citenamefont {Bar-On}, \citenamefont {Phillips},\ and\ \citenamefont
  {Milo}}]{shamir2016snapshot}%
  \BibitemOpen
  \bibfield  {author} {\bibinfo {author} {\bibfnamefont {M.}~\bibnamefont
  {Shamir}}, \bibinfo {author} {\bibfnamefont {Y.}~\bibnamefont {Bar-On}},
  \bibinfo {author} {\bibfnamefont {R.}~\bibnamefont {Phillips}},\ and\
  \bibinfo {author} {\bibfnamefont {R.}~\bibnamefont {Milo}},\ }\bibfield
  {title} {\bibinfo {title} {Snapshot: timescales in cell biology},\
  }\href@noop {} {\bibfield  {journal} {\bibinfo  {journal} {Cell}\ }\textbf
  {\bibinfo {volume} {164}},\ \bibinfo {pages} {1302} (\bibinfo {year}
  {2016})}\BibitemShut {NoStop}%
\bibitem [{\citenamefont {Chen}\ \emph {et~al.}(2014)\citenamefont {Chen},
  \citenamefont {Li}, \citenamefont {Hsu}, \citenamefont {Zhang}, \citenamefont
  {Lai}, \citenamefont {Chan},\ and\ \citenamefont {Chen}}]{Chen2014}%
  \BibitemOpen
  \bibfield  {author} {\bibinfo {author} {\bibfnamefont {S.}~\bibnamefont
  {Chen}}, \bibinfo {author} {\bibfnamefont {N.}~\bibnamefont {Li}}, \bibinfo
  {author} {\bibfnamefont {S.~F.}\ \bibnamefont {Hsu}}, \bibinfo {author}
  {\bibfnamefont {J.}~\bibnamefont {Zhang}}, \bibinfo {author} {\bibfnamefont
  {P.~Y.}\ \bibnamefont {Lai}}, \bibinfo {author} {\bibfnamefont {C.~K.}\
  \bibnamefont {Chan}},\ and\ \bibinfo {author} {\bibfnamefont
  {W.}~\bibnamefont {Chen}},\ }\bibfield  {title} {\bibinfo {title} {Intrinsic
  fluctuations of cell migration under different cellular densities},\
  }\bibfield  {journal} {\bibinfo  {journal} {Soft Matter}\ }\textbf {\bibinfo
  {volume} {10}},\ \href {https://doi.org/10.1039/c3sm52752b}
  {10.1039/c3sm52752b} (\bibinfo {year} {2014})\BibitemShut {NoStop}%
\bibitem [{\citenamefont {Caballero}\ \emph {et~al.}(2014)\citenamefont
  {Caballero}, \citenamefont {Voituriez},\ and\ \citenamefont
  {Riveline}}]{Caballero2014}%
  \BibitemOpen
  \bibfield  {author} {\bibinfo {author} {\bibfnamefont {D.}~\bibnamefont
  {Caballero}}, \bibinfo {author} {\bibfnamefont {R.}~\bibnamefont
  {Voituriez}},\ and\ \bibinfo {author} {\bibfnamefont {D.}~\bibnamefont
  {Riveline}},\ }\bibfield  {title} {\bibinfo {title} {Protrusion fluctuations
  direct cell motion},\ }\bibfield  {journal} {\bibinfo  {journal} {Biophysical
  Journal}\ }\textbf {\bibinfo {volume} {107}},\ \href
  {https://doi.org/10.1016/j.bpj.2014.05.002} {10.1016/j.bpj.2014.05.002}
  (\bibinfo {year} {2014})\BibitemShut {NoStop}%
\bibitem [{\citenamefont {Ester}\ \emph {et~al.}(1996)\citenamefont {Ester},
  \citenamefont {Kriegel}, \citenamefont {Sander},\ and\ \citenamefont
  {Xu}}]{Ester1996}%
  \BibitemOpen
  \bibfield  {author} {\bibinfo {author} {\bibfnamefont {M.}~\bibnamefont
  {Ester}}, \bibinfo {author} {\bibfnamefont {H.-P.}\ \bibnamefont {Kriegel}},
  \bibinfo {author} {\bibfnamefont {J.}~\bibnamefont {Sander}},\ and\ \bibinfo
  {author} {\bibfnamefont {X.}~\bibnamefont {Xu}},\ }\bibfield  {title}
  {\bibinfo {title} {A density-based algorithm for discovering clusters in
  large spatial databases with noise},\ }\href@noop {} {\bibfield  {journal}
  {\bibinfo  {journal} {KDD Proceedings, Proceedings of Second International
  Conference on Knowledge Discovery and Data Mining}\ }\textbf {\bibinfo
  {volume} {96}} (\bibinfo {year} {1996})}\BibitemShut {NoStop}%
\bibitem [{\citenamefont {Inc.}(2021)}]{MATLAB}%
  \BibitemOpen
  \bibfield  {author} {\bibinfo {author} {\bibfnamefont {T.~M.}\ \bibnamefont
  {Inc.}},\ }\href {https://www.mathworks.com/help/stats/dbscan.html} {\bibinfo
  {title} {Matlab version: 9.11.0.1837725 (r2021b)}} (\bibinfo {year}
  {2021})\BibitemShut {NoStop}%
\bibitem [{\citenamefont {Lintz}\ \emph {et~al.}(2017)\citenamefont {Lintz},
  \citenamefont {Muñoz},\ and\ \citenamefont {Reinhart-King}}]{Lintz2017}%
  \BibitemOpen
  \bibfield  {author} {\bibinfo {author} {\bibfnamefont {M.}~\bibnamefont
  {Lintz}}, \bibinfo {author} {\bibfnamefont {A.}~\bibnamefont {Muñoz}},\ and\
  \bibinfo {author} {\bibfnamefont {C.~A.}\ \bibnamefont {Reinhart-King}},\
  }\bibfield  {title} {\bibinfo {title} {The mechanics of single cell and
  collective migration of tumor cells},\ }\bibfield  {journal} {\bibinfo
  {journal} {Journal of Biomechanical Engineering}\ }\textbf {\bibinfo {volume}
  {139}},\ \href {https://doi.org/10.1115/1.4035121} {10.1115/1.4035121}
  (\bibinfo {year} {2017})\BibitemShut {NoStop}%
\bibitem [{\citenamefont {Peruani}\ \emph {et~al.}(2011)\citenamefont
  {Peruani}, \citenamefont {Ginelli}, \citenamefont {Bär},\ and\ \citenamefont
  {Chaté}}]{Peruani2011}%
  \BibitemOpen
  \bibfield  {author} {\bibinfo {author} {\bibfnamefont {F.}~\bibnamefont
  {Peruani}}, \bibinfo {author} {\bibfnamefont {F.}~\bibnamefont {Ginelli}},
  \bibinfo {author} {\bibfnamefont {M.}~\bibnamefont {Bär}},\ and\ \bibinfo
  {author} {\bibfnamefont {H.}~\bibnamefont {Chaté}},\ }\bibfield  {title}
  {\bibinfo {title} {Polar vs. apolar alignment in systems of polar
  self-propelled particles},\ }\bibfield  {journal} {\bibinfo  {journal}
  {Journal of Physics: Conference Series}\ }\textbf {\bibinfo {volume} {297}},\
  \href {https://doi.org/10.1088/1742-6596/297/1/012014}
  {10.1088/1742-6596/297/1/012014} (\bibinfo {year} {2011})\BibitemShut
  {NoStop}%
\bibitem [{\citenamefont {Bisson}(2021)}]{Bisson2021}%
  \BibitemOpen
  \bibfield  {author} {\bibinfo {author} {\bibfnamefont {A.}~\bibnamefont
  {Bisson}},\ }\bibfield  {title} {\bibinfo {title} {Preprint highlight:
  Pressure and curvature control of contact inhibition in epithelia growing
  under spherical confinement},\ }\bibfield  {journal} {\bibinfo  {journal}
  {Molecular biology of the cell}\ }\textbf {\bibinfo {volume} {32}},\ \href
  {https://doi.org/10.1091/mbc.E21-10-0124p} {10.1091/mbc.E21-10-0124p}
  (\bibinfo {year} {2021})\BibitemShut {NoStop}%
\end{thebibliography}%
\end{document}